\documentclass[graybox, nosecnum]{svmult}

\usepackage{mathptmx}       
\usepackage{helvet}         
\usepackage{courier}        
\usepackage{type1cm}        
\usepackage{makeidx}         
\usepackage{graphicx}        
\usepackage{multicol}        
\usepackage[bottom]{footmisc}
\usepackage{hyperref}        
\usepackage{soul}            
\hypersetup{colorlinks=true,urlcolor=blue}
\usepackage[square,numbers]{natbib}
\makeindex             

\def\gtsima{$\; \buildrel > \over \sim \;$}
\def\ltsima{$\; \buildrel < \over \sim \;$}
\def\gsim{\lower.5ex\hbox{\gtsima}}
\def\lsim{\lower.5ex\hbox{\ltsima}}

\begin{document}
\title*{Probing black-hole accretion through time variability}
\author{Barbara De Marco \thanks{corresponding author}, Sara E. Motta and Tomaso M. Belloni}
\institute{B. De Marco \at Departament de F\`isica, EEBE, Universitat Polit\`ecnica de Catalunya, Av. Eduard Maristany 16, 08019 Barcelona, Spain, \email{barbara.de.marco@upc.edu}
\and S.E. Motta \at INAF - Osservatorio Astronomico di Brera, via E. Bianchi 46, 23087, Merate, Italy \email{sara.motta@inaf.it}
\and T.M. Belloni \at INAF - Osservatorio Astronomico di Brera, via E. Bianchi 46, 23087, Merate, Italy \email{tomaso.belloni@inaf.it}}
%
%
\maketitle
\abstract{Flux variability is a remarkable property of black hole (BH) accreting systems, and a powerful tool to investigate the multi-scale structure of the accretion flow. The X-ray band is where some of the most rapid variations occur, pointing to an origin in the innermost regions close to the BH. The study of fast time variability provides us with means to explore the accretion flow around compact objects in ways which are inaccessible via spectral analysis alone, and to peek at regions which cannot be imaged with the currently available instrumentation.
In this chapter we will discuss fast X-ray variability in stellar-mass BH systems, namely binary systems containing a star and a BH, occasionally contrasting it with observations of supermassive BHs in active galactic nuclei. We will explore how rapid variations of the X-ray flux have been used in multiple studies as a diagnostic of the innermost regions of the accretion flow in these systems.
To this aim we will provide an overview of the currently most used analysis approaches for the study of X-ray variability, describe observations of both aperiodic and quasi-periodic phenomena, and discuss some of the proposed models. 
}
\section{Keywords} 
X-rays: binaries -- accretion, accretion discs -- stars: black holes

\section{Introduction}
One of the main ways to detect black holes (BH) is by their gravitational influence on the surrounding gas. In particular, BHs can be (almost) unambiguously identified when the system is in an ``active state'', during which the accretion of gas onto the BH proceeds at relatively high rates. For stellar-mass ($M_{\rm{BH}}\sim 10\  M_{\odot}$) BHs in a binary system this occurs when significant amounts of gas from a low mass ($M_2<1\ M_{\odot}$) or a high mass (typically $M_2\gsim M_{\rm{BH}}$) companion star are transferred to the BH, respectively via Roche-Lobe overflow or stellar winds.
The process leads to the accreting material being heated up to high temperatures, eventually becoming luminous in the X-rays \citep[][]{Lasota2001}.
Because of their brightness during these active phases and their rapid excursions into different accretion states, BH X-ray binaries (XRBs) represent ideal laboratories to study the condition of matter in the vicinity of the BH, and offer us the unique opportunity to witness in a single source and over ``humanly accessible'' time scales many (if not all) the flavours that mark out the process of accretion onto a BH.

\begin{figure}
\centering
\includegraphics[width=1.0\textwidth]{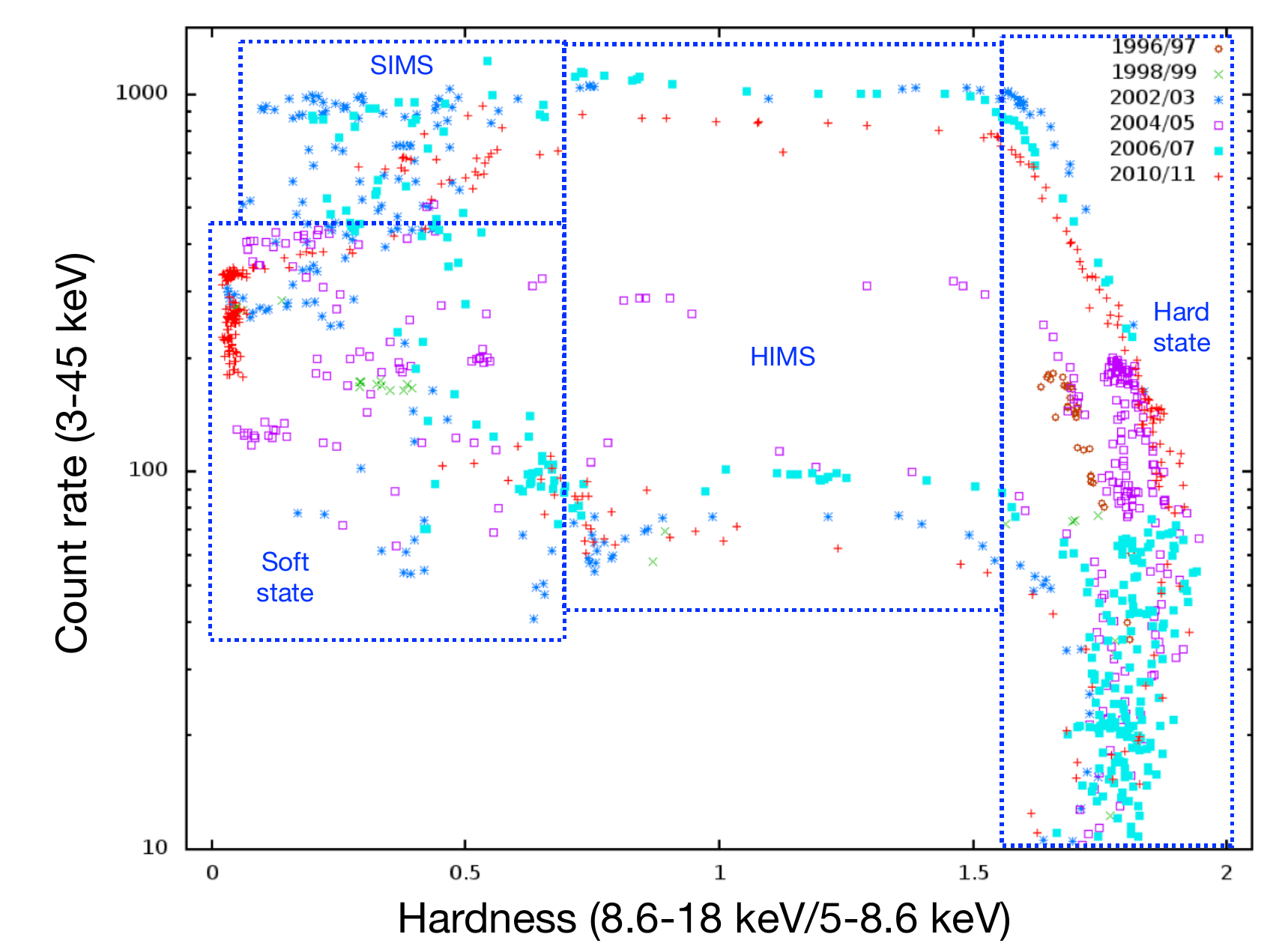}
\caption{The days-to-months luminosity vs. hardness evolution (hardness-intensity diagram, HID) of the transient BH XRB system GX 339-4, as observed by {\it RXTE} during 6 different outbursts (years reported in the upper-right label). Adapted from \cite{Dzielak2019} using the schematic description of the different accretion states reported in \cite{Belloni2005a}.} \label{fig:HID}
\end{figure}

X-ray flux variations are a characteristic feature of BH XRBs. The observed variability is distributed over a wide range of time scales, with the long-term variations (on time scales of days-months) probing changes in the global structure of the accretion flow, and the fastest variations (on time scales shorter than thousands of seconds down to milliseconds) probing the smallest scales in the vicinity of the BH.\\
The days-to-months variability of BH XRBs is most commonly described by means of combined changes in their spectral hardness and luminosity (Fig. \ref{fig:HID}), this representation being much more coherent than a simple flux vs. time (a.k.a light curve) description, see \cite{Belloni2010a, Belloni2016}.
The onset of an outburst is determined by a rapid rise of mass transfer above the level characterising the quiescent state (which is of the order of $\sim10^{-11}M_{\odot}\rm{yr}^{-1}$, \citep[e.g.][]{Narayan1996}, with typical X-ray luminosities of $L_{X}\sim10^{30-34} \rm{erg/s}$), in which the great majority of BH XRBs spends most of their lives. Such a rise of mass transfer rate is thought to happen as a consequence of a viscous instability \citep[e.g.][]{Lasota2001,Dubus2001} triggered by the increase of gas temperature (above the threshold for the ionisation of hydrogen) in an initially truncated, geometrically thin accretion disc \citep[e.g.][]{Bernardini2016}. 
These systems manifest themselves as \emph{transients} due to their recurrent and relatively short-duration outbursts, culminating in an extraordinary increase of X-ray luminosity (typically in excess of $L_{X}\sim 10^{36-38}\rm{erg/sec}$), and terminating with some event (e.g. a second instability or the depletion of the outer disc by a wind, \cite{MunozDarias2016}), which brings the disc back to a cool neutral state.
On the other side are a few known \emph{persistent} BH XRBs, which are always active or can remain active for several decades (Cyg X-1 and GRS 1915+105 are two examples of known persistent sources).\\
Both transient and persistent BH XRBs undergo rapid transitions between different accretion states \citep[e.g.][]{Belloni2010a,Dunn2010,Belloni2016}. The main (in terms of the time the source spends in these states) accretion states are the so-called \textbf{hard state} (HS) and \textbf{soft state} (SS). The hard state is spectrally characterised by the dominance of a Comptonised hard X-ray (power-law) component, with photon index $\sim 1.7$ and peak emission at $\sim60-100$ keV \citep[e.g.][]{Joinet2008,Motta2009a,Gilfanov2010}, produced in an inner hot flow or corona. The soft state is instead dominated by a soft thermal disc (multi-temperature black body) component, with an inner disc  temperature of $\sim$ 0.5-1 keV \citep[e.g.][]{Gierlinski2004}. 
In between are a number of (short-lived) intermediate states, characterised by a significant softening of the spectrum due to the concurrent presence of a strong disc component and a hard power law with photon index increasing up to $\Gamma\sim2.0-2.5$. 
On the one hand, persistent sources are observed to transition back and forth between hard and soft states, through a number of intermediate states, at almost constant luminosity\footnote{Deviations from the described fundamental behaviour, which involve occasional excursions to high luminosity steep power law states, dubbed ``anomalous states'' (\cite{Belloni2010a}) or \textbf{ultraluminous states} (ULS; \citep[e.g.][]{Done2004, Motta2009a}), can be observed in both persistent and transient sources.}. On the other hand, a transient source always goes through the same hysteresis cycle, commonly represented in a ``Hardness-Intensity diagram'' (HID, see Fig. \ref{fig:HID}, where the hardness refers to the ratio between fluxes in a hard and a soft X-ray band) \citep[e.g.][]{Belloni2010a,Belloni2016}: starting from a low luminosity hard state it undergoes a sharp increase (by about three-four orders of magnitude) of X-ray luminosity with mild changes of spectral hardness, reaching the maximum luminosity over a period of a few weeks to a few  months; then it rapidly (a few days) transitions to a \textbf{hard-intermediate state} (HIMS) followed by a \textbf{soft-intermediate state} (SIMS) , both characterised by strong spectral changes at almost constant luminosity; it finally reaches the soft state, where it can remain for months with fluctuations in luminosity of about one-two orders of magnitude; at the end of the outburst the luminosity drops, and the source transitions back to the hard state, where the luminosity further decreases eventually leading the source back to quiescence.

The most rapid (thousands of seconds to well under a second) X-ray variability in BH XRBs is strongly correlated with the spectral properties of the source, and therefore its accretion states \citep[e.g.][]{Belloni2011}.
 This fast variability cannot be easily appreciated by simply looking at changes in spectral hardness and luminosity of the source, and the inspection of a source's light curve may be unpractical to useless depending on how fast the signal of interest is. Hence, the sub-minute time variability in BH XRBs is best studied in the Fourier domain, and the power density spectrum (PDS, or power spectrum; see \cite{Press1992} and the chapter "Basics of Fourier Analysis for High-Energy Astronomy" in this book) is to date still the most common tool to study this rapidly variable emission. The study of the PDS of BH XRBs revealed the most rapid X-ray variability in these sources to be a combination of aperiodic variability components (i.e., scale-free broad-band noise components, covering several decades in frequency, \citep[e.g.][]{Belloni2005a,Belloni2010}, which account for most of the observed X-ray variability in the large majority of BH XRB accretion states.  Quasi-periodic oscillations (QPOs), narrow peaks of width less than $\sim1/2$ the centroid frequency \citep[e.g.][]{Ingram2019},  typically appear on top of the aperiodic variability during high luminosity hard and intermediate states, but can be the dominant component during the intermediate states.
 
In this chapter we will focus on discussing the complex phenomenology that characterises the fast X-ray variability in BH XRBs.

\section{X-ray variability in BH XRBs}

The distribution of variability power as a function of temporal frequency (i.e. the inverse of the time scale of variability) is obtained by computing the PDS, namely the modulus squared of the Fourier transform of the light curve (i.e. $P(\nu)=A |X(\nu)|^2$, where $X(\nu)$ is the Fourier transform of the light curve $x(t)$, and $A$ is a normalisation constant; see \cite{Press1992} and the chapter "Basics of Fourier Analysis for High-Energy Astronomy" in this book). The study of the PDS enables characteristic variability time scales and different variability components to be easily found in the data. Moreover, the integral of the PDS over a given interval of temporal frequencies ($[\nu_1:\nu_2]$) yields the contribution to the observed variability due to variations between the corresponding time scales ($[1/\nu_1:1/\nu_2]$), i.e. the root mean square (rms) variability amplitude. The rms allows quantifying the level of intrinsic variability of the source (once the contribution from counting noise is subtracted out)  \citep[][]{Vaughan2003}. When normalised by the mean count rate, the rms is expressed in fractional units (fractional rms, $F_{\rm var}$; \citep[][]{Belloni1990} and the chapter "Basics of Fourier Analysis for High-Energy Astronomy" in this book).

Early studies of the X-ray PDS of outbursting BH XRBs in the hard state revealed strong variability (fractional rms of tens of percent), which could be described in terms of a continuous broad-band noise component, with no obvious features save from a few breaks, possibly ascribed to  some characteristic time scales of the system \citep[e.g.][]{Nolan_1981a,Belloni1990a}. The later advent of all-sky monitors coupled with large effective area instruments greatly boosted the study of X-ray variability in BH XRBs, and enabled the prompt detection of transient sources and intensive follow-up. As a result, the number of known BH XRBs has been continuously growing \citep[][]{Corral-Santana2016,Tetarenko2016}. The new studies unveiled complex structures in the PDS of BH XRBs, previously hidden by the poorer signal-to-noise of the data. In particular, first the instruments on-board the Ginga satellite and then the Rossi X-ray Timing Explorer ({\it RXTE}) greatly enriched our archives, which became (and still are) goldmines of data for the investigation of the X-ray timing properties of BH XRBs. This resulted in the discovery of new rich phenomenology, including various types of QPOs and peaked noise components, which brought to a thorough classification of timing features and increasingly more ambitious models to be developed.

\subsection{Time scales of variability}
Before delving into the description of X-ray variability features in BH XRBs, it is important to discuss what kind of time scales we expect to observe. 
Indeed, theoretical models define characteristic time scales over which the structure of the accretion flow can vary \cite{Frank2002}. These time scales depend on the parameters of the BH and of the accretion flow (e.g. the BH mass, the viscosity parameter $\alpha$, the scale height $H/R$ of the flow), as well as on the emitting radius \citep[e.g.][]{Peterson2001}. 
In the inner regions of the accretion flow, some of these time scales can be rather short and under certain circumstances (in particular when specific radii are contributing more to the emission) can clearly show up in the data. At each radius, the shortest possible characteristic time scale is the dynamical time scale, which corresponds to the inverse of the orbital (Keplerian) frequency, and the time scale at which hydrostatic equilibrium is restored in the vertical direction \cite{Frank2002}.
Other important time scales are the viscous time scale, over which matter diffuses through the disc due to viscous torques, and the thermal time scale, over which thermal equilibrium is restored. The thermal and dynamical time scales are rather fast (the dynamical one can reach milliseconds in the case of small radii around neutron stars, NS), while the viscous time scale is considerably slower. 
In addition, one can consider the free-fall time scale and the sound crossing time scale. 
With the exception of the dynamical and free-fall time scales, all these time scales are connected to the accretion flow and would make no sense in presence of a single orbiting particle. In addition to the orbital frequency, General Relativity predicts two epicyclic frequencies associated with a given orbit: the radial frequency, which is associated with changes in radius in case of an elliptical orbit, and the vertical frequency, which arises when small perturbations tilt an orbit away from the plane perpendicular to the spin axis of the central object. 

All these time scales (except the epicyclic frequencies in Kerr spacetime, which are non-monotonous near ISCO) increase with increasing distance from the compact object, but within several gravitational radii they are shorter than one second.
The problem of identifying the frequency of a detected temporal feature with one of these time scales is very complex, due to the dependence on many unknown parameters of the accretion flow. Only multiple detections, whether simultaneous or in different observations, can help this identification.

\subsection{Aperiodic X-ray variability}

\begin{figure}
\centering
\includegraphics[width=0.32\textwidth]{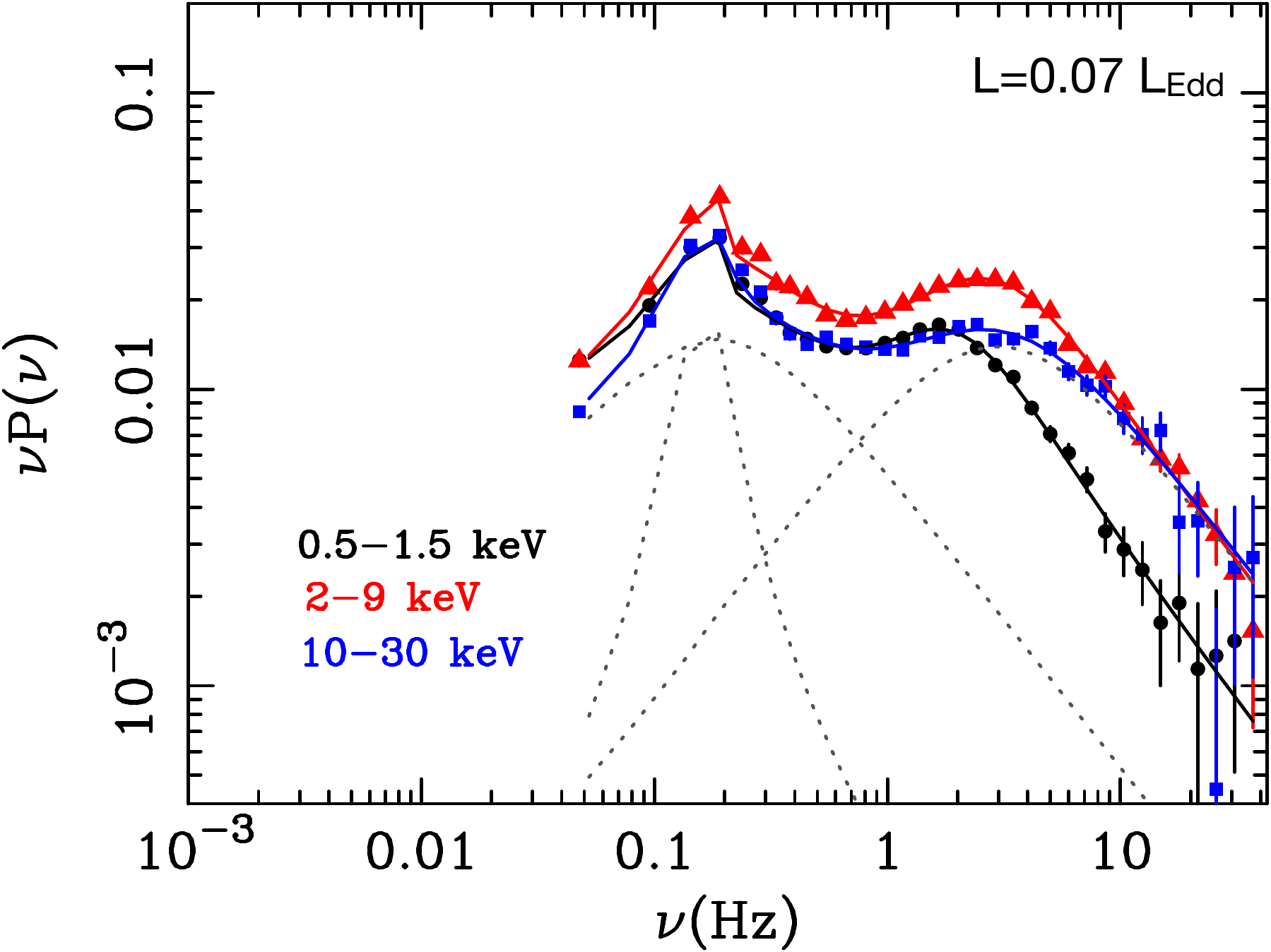}
\includegraphics[width=0.3\textwidth]{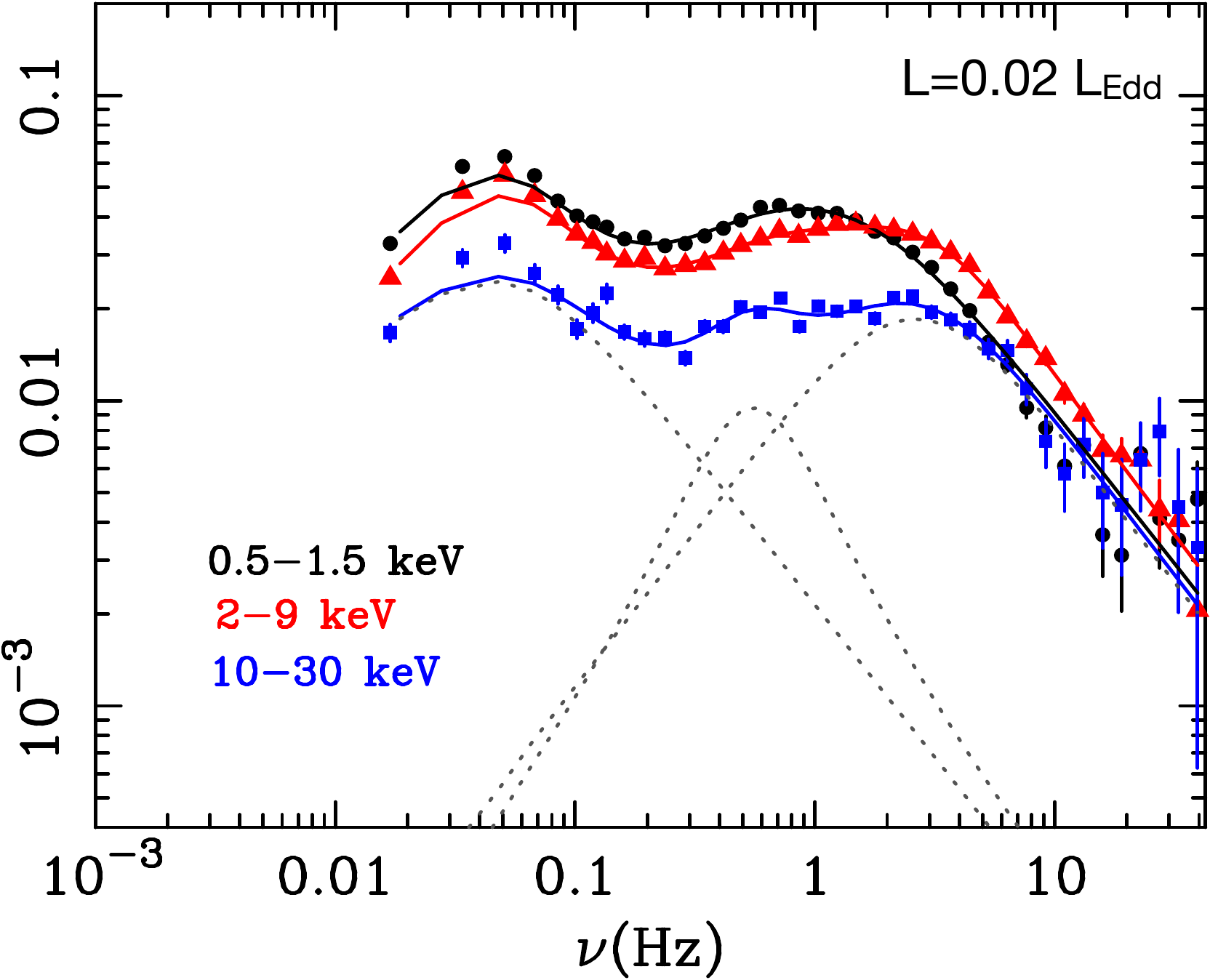}
\includegraphics[width=0.3\textwidth]{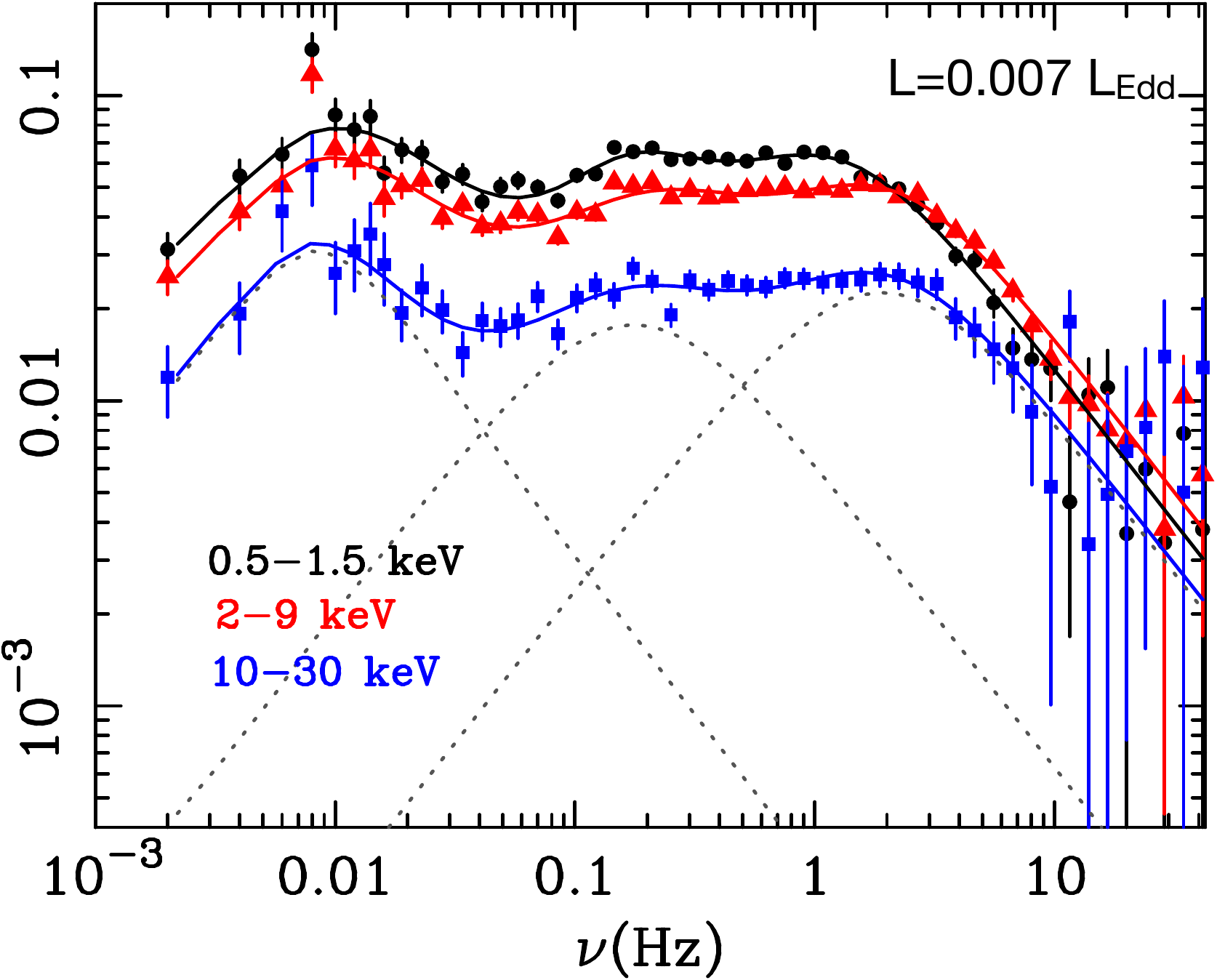}
\caption{The evolution of the PDS in GX 339-4 in the energy bands 0.5-1.5 keV (XMM-Newton), 2-9 keV (XMM-Newton), and 10-30 keV ({\it RXTE}). Each plot refers to different values of hard X-ray (3-30 keV) luminosity in the hard state. The PDSs are fit with a sum of Lorentzians, the dotted lines are the best-fit set of Lorentzians for the 10-30 keV PDS. Adapted from \cite{DeMarco2015a}. }\label{fig:PDS_GX}
\end{figure}

As highlighted in the introduction to this chapter, aperiodic broad-band noise is the major contributor to the X-ray variability of BH XRBs. This aperiodic variability is not smoothly distributed, but shows enhanced power around specific frequencies/time scales, resulting in a ``bumpy'' PDS. This peculiar shape emerges from the combination of several broad features, which appear as peaked noise components in the $\nu P(\nu)$ representation of the PDS \cite{Belloni2002} (Fig. \ref{fig:PDS_GX}). Each noise component in the PDS may be the signature of turbulence-induced variations of the mass accretion rate occurring over a given time scale \citep[e.g.][]{Hogg2016,Bollimpalli2020}, coupled with enhanced dissipation in specific regions of the accretion flow \cite{Mahmoud2018}, thus resulting in a more prominent contribution from variability produced at certain radii.
A peaked noise PDS may also result from interference of the variable X-ray emitted signals produced in distinct regions of the accretion flow (e.g. the inner and outer radii of a radially extended Comptonising region), with a delay between them corresponding to the time needed for the signal to propagate from one region to the other \cite{Veledina2016}.

A firm identification of these theoretical time scales in the PDS has the potential to provide constraints on the parameters of the accretion flow and of the BH itself  \citep[e.g.][]{Gilfanov2005,Motta2014}. 
However, the highly complex structure of the PDS significantly challenges the modelling of X-ray variability in BH XRBs. Therefore, a solid link between the observed time scales of X-ray variability identified in the PDS, and the theoretical variability time scales predicted by accretion disc models has proven  hard to establish. 
Nonetheless, we can still obtain a qualitative idea of the ways the accretion flow may evolve as the source switches between different accretion modes by monitoring the changes in the PDS during an outburst. 
The most convenient and effective way to track these changes consists in describing the multi-peaked noise shape of the PDS as a superposition of variability components empirically represented by Lorentzians\footnote{From a physical point of view a Lorentzian component in the PDS could result from a dampened oscillator.} of different widths \cite{Belloni2002} (Fig. \ref{fig:PDS_GX}). In this approach, broad Lorentzians well describe the aperiodic variability components, while narrow Lorentzians are used to represent quasi periodic features, thus allowing variability components of possibly different origin to be described under the same formalism, and their relationship to be easily studied. 

Analysis of the PDS of BH XRBs reveals that each X-ray variability component significantly evolves throughout the outburst (see Fig. \ref{fig:PDS_GX}), following tight correlations \citep[e.g.][]{Psaltis1999,Wijnands1999a}. 
In particular, broad components are observed to vary in normalisation, width, and centroid temporal frequency.
Such behaviour has been interpreted as the signature of major variations in the structure of the innermost accretion flow. In particular, the peak frequency of each noise component \footnote{The peaks of variability power correspond to the frequency $\nu_{max}=\sqrt{\Delta^2+\nu_0^2}$, where $\nu_0$ is the centroid temporal frequency, and $\Delta$ the half-width at half maximum of the Lorentzian, see \cite{Belloni2002}} in the PDS is observed to steadily shift towards higher frequencies \citep[e.g.][]{Belloni2005a} as the source evolves throughout the hard state and transitions to the soft state (the opposite trend is observed when the source leaves the soft state and returns to the hard state). 
This behaviour is characteristic of both the  aperiodic and quasi-periodic components  of the PDS (see next section), and leads to tight correlations that may change across the different accretion states \citep[e.g.][]{Psaltis1999,Wijnands1999a}, suggesting a scaling as a function of a common physical parameter. 
Given the dependence of all physical time scales (see previous section) on the emitting radius, such trends give support to models that predict a gradual decrease of characteristic radii of the accretion flow as the source moves from the hard to the soft state \citep[e.g.][]{Esin1997,Meyer2000,Ferreira2006}.
Such a shift occurs up to a maximum temporal frequency (of the order $\sim 10$ Hz), above which the PDS shows a drop of aperiodic X-ray variability power, likely marking the presence of a minimum radius in the hot inner flow. For example, assuming that such variability is the signature of viscous processes in a geometrically thick hot flow (with thickness $H/R\sim0.3$), and assuming a value of $\alpha\sim0.5$ for the viscosity parameter \citep[e.g.][]{Tetarenko2018}, this minimum radius would approximately coincide with the last stable circular orbit around a BH of $10\ {\rm M_{\odot}}$.

\subsection{Quasi periodic oscillations}

QPOs have been discovered in the late '70s \cite{Samimi1979}, 
when the reference to a `sporadic quasi-periodic behaviour' of the X-ray light curve of the BH XRB GX 339-4, as observed by \textit{HEAO}, appeared for the first time in the literature. 
QPOs have been observed since then in virtually all BH and NS X-ray binaries, and detections have been also reported in some ultraluminous X-ray sources (ULX) and active galactic nuclei (AGN) \cite{Kaaret2017, Alston2016}.
The first robust detection of a QPO in a binary system was reported by \citet{Motch1983}, based on data of GX 339-4 from the \emph{Ariel 6} rocket. QPOs were later found in the X-ray data of the NS source GX 5-1 as observed by \emph{EXOSAT} \cite{vanderKlis1985}. In the following years, the repeated detection of QPOs in the data from the \emph{EXOSAT} and \emph{GINGA} satellites highlighted the existence of several different types of QPOs in both NS \cite{Middleditch1986,VdK1989} and BH XRBs \cite{Miyamoto1991a}.
The richest database for the study of QPOs in XRBs is to date the one accumulated by the {\it RXTE} satellite, which observed virtually all the known (up to the year 2012) BH and NS XRBs, detecting hundreds of QPOs. 

QPOs reflect the oscillations of the accretion flow in the presence of a strong gravitational field, possibly excited in narrow ranges of radii. They have centroid frequencies that can be accurately measured, and can be associated with accretion-related time scales. Being produced in the vicinity of relativistic objects such as BHs and NSs, QPOs are expected to carry information about the properties and motion of matter in the strong field regime, thus providing powerful probes of the predictions of the theory of General Relativity. Moreover, the observation of QPOs in specific accretion states and during transitions indicates that they could be a key ingredient to understand the physical processes behind the observed phenomenology.

In BH systems, QPOs are generally divided into two main groups: the low frequency (LF) QPOs, and the high frequency (HF) QPOs. The former show a centroid frequency below $\sim 30$ Hz, while the latter are generally observed above a few hundreds Hz \cite{Belloni2012}. It is worth noting that a similar distinction applies to QPOs observed in NS systems, although the nomenclature is more complex in that case, with extra QPO classes, and other small differences mainly in the QPO naming (e.g., the highest frequency QPOs in NSs are generally referred to as ``kHz QPOs'' rather than ``HF QPOs'').

\subsubsection{Low-frequency quasi periodic oscillations}

LF QPOs are commonly observed in virtually all BH XRBs with high intrinsic amplitudes \cite{Motta2015}, which has allowed a detailed picture of their properties to be drawn.
LF QPOs in BH XRBs have traditionally been classified into three types: A, B and C. Type-C QPOs are mostly observed in the hard state, but they are often detected also in the HIMS, and (less often) in the soft state. Instead, Type-B and Type-A QPOs are only seen across the transition to the soft state.

\begin{figure}
\centering
\includegraphics[width=0.45\textwidth]{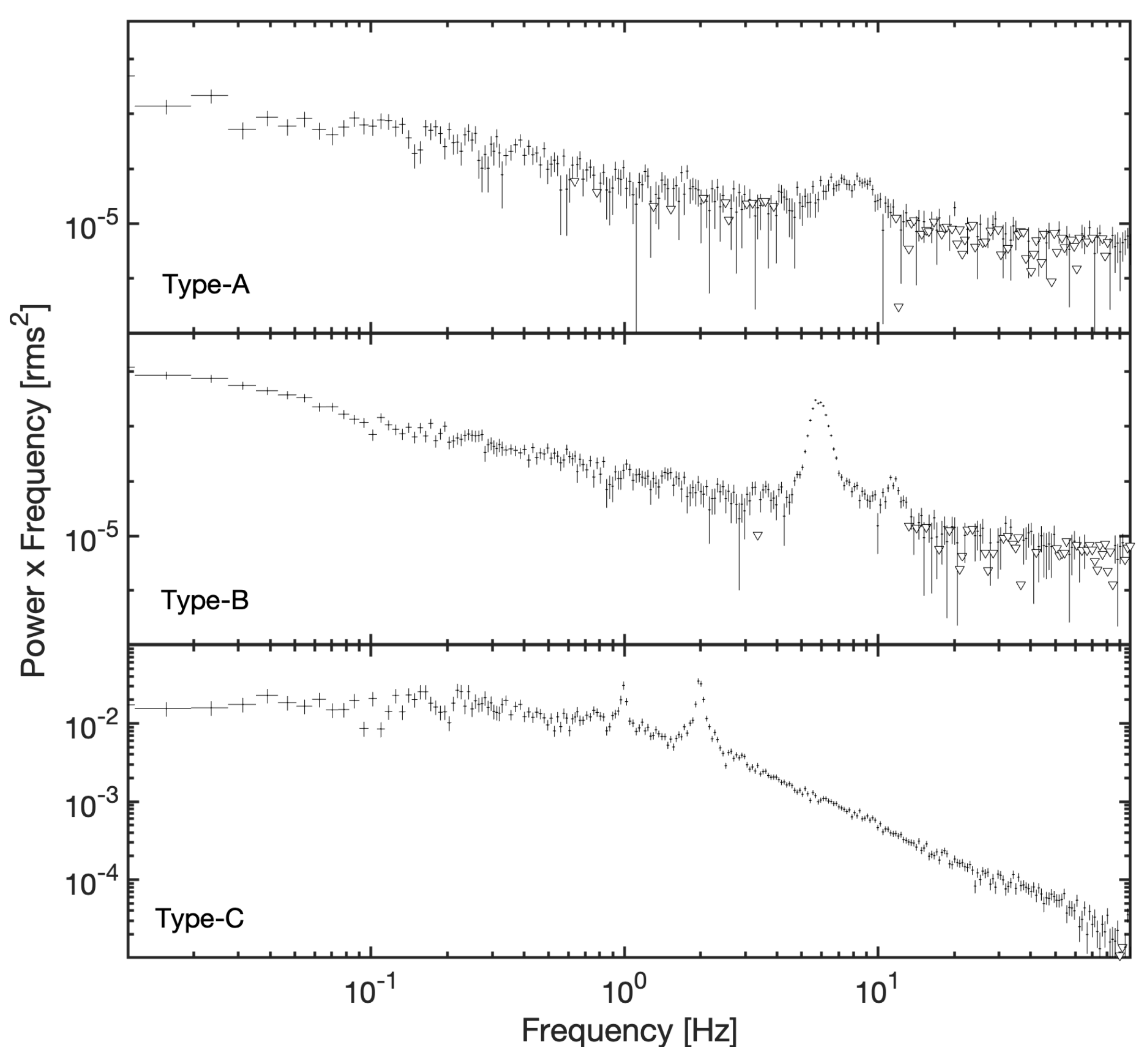}
\includegraphics[width=0.45\textwidth]{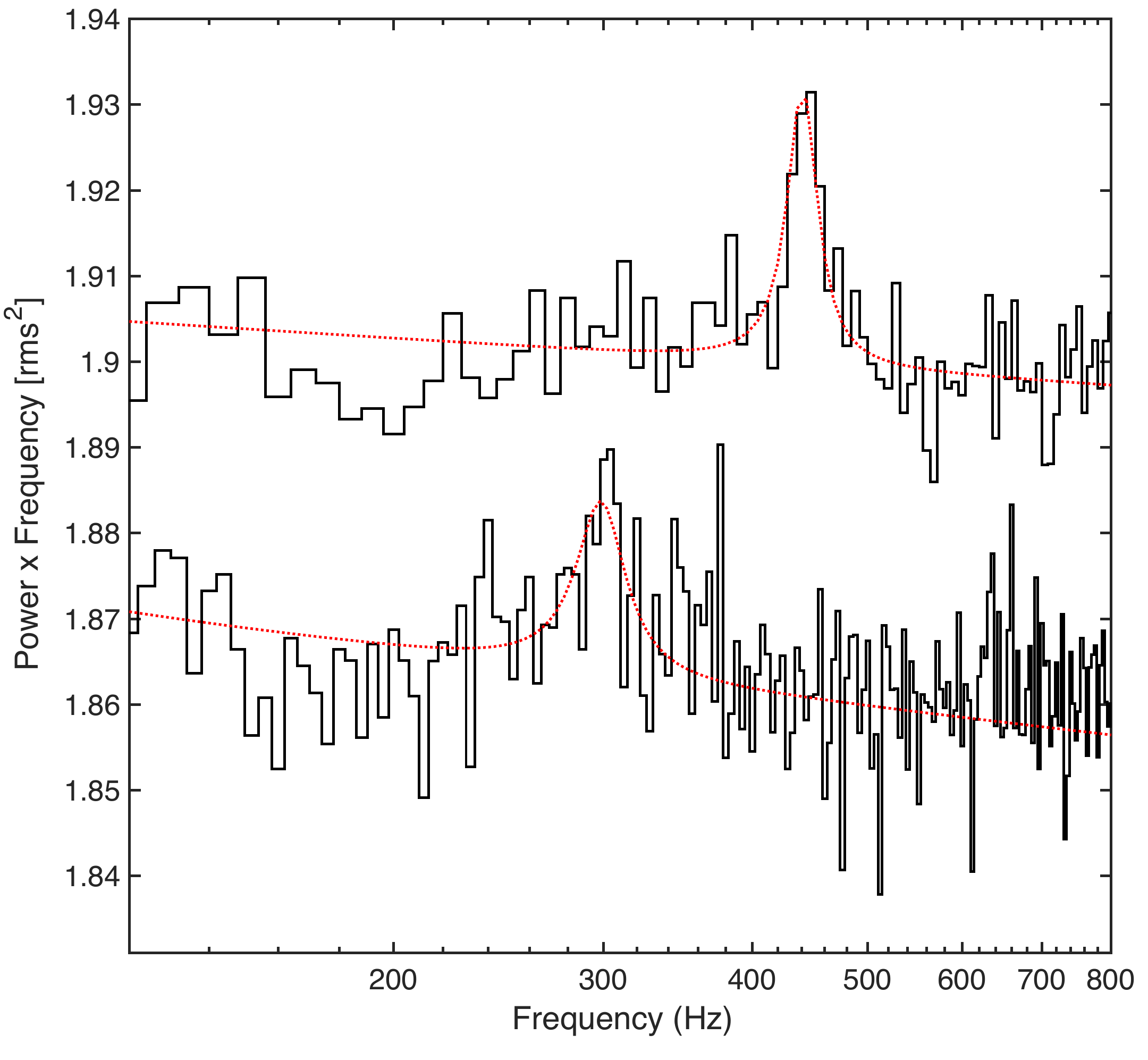}
\caption{Left panel: the three classes of low-frequency QPOs observed in BH XRBs. Right panel: the HFQPOs pair observed in the black hole binary GRO J1655-40. The QPO at $\sim$500 Hz (lower PDS) is detected in a harder energy band than that at $\sim$350 Hz (upper PDS). The upper PDS has been shifted upward for clarity. The red line marks the best fitting  model. Uncertainties are not plotted for clarity. Figures adapted from \cite{Motta2011} and \cite{Motta2012}. }\label{fig:QPOs}
\end{figure}

\textbf{Type-C QPOs}  are by far the most common type of QPOs in BH systems. Type-C QPOs can be detected in all accretion states (including the ULS, except the SIMS), although they are most prominent in the HIMS and at the bright end of the hard state. The centroid frequency of type-C QPOs is highly sensitive to the state of the source, rising from a few mHz in the hard state at low luminosities, to $\sim$10 Hz in the HIMS, and up to $\sim$30 Hz in the ULS. 
These QPOs are characterised by a high-amplitude (up to $F_{\rm var} \sim20\%$) and a narrow peak (Q $>2$, where Q is the ratio between the centroid frequency and the width of the QPO peak, also called the \textit{quality factor}) in the power spectrum, superimposed to  a ‘flat-top’ broad band noise characterised by a low and high frequency break. A number of harmonics are generally detected, of which the one with the highest rms amplitude is usually (but not exclusively) identified with the fundamental.  The frequency of the Type-C QPO correlates with the low frequency break in the PDS \cite{Wijnands1999a}, as well as with a number of spectral parameters, such as the photon index, the disc temperature and the disc truncation radius (see, e.g., \cite{Muno1999, rodrigue02_aei,Vignarca2003, Motta2011}). Type-C QPOs have also been observed at optical (e.g. \cite{Motch1983,Imamura1990,Gandhi2010}), ultraviolet \cite{Hynes2003a} and infrared \cite{Kalamkar2016,Vincentelli2019} wavelengths. Simultaneous multi-wavelength observations have revealed that the UV/optical/IR QPO centroid frequency is sometimes consistent with the X-ray QPO fundamental frequency \cite{Hynes2003a,Durant2009,Gandhi2010} and sometimes with half of the X-ray fundamental frequency \cite{Motch1983,Kalamkar2016}.\\
\textbf{Type-B QPOs}  have been detected in a large number of BH XRBs and they appear during the SIMS, which is in fact \textit{defined} by the presence of a Type-B QPO \cite{Belloni2016} and a weak red noise variability component (usually modelled with a zero-centered broad Lorentzian, or a power law). They are characterised by a relatively high amplitude (up to $F_{\rm var}\sim$5\% ) and narrow ($Q \gsim 6$) peak, and generally show centroid frequency at 5-6 Hz (even though type-B QPOs with significantly lower centroid frequencies have been also reported \cite{Motta2011}). A weak second harmonic is often present, sometimes together with a sub-harmonic peak. In a few cases, the sub-harmonic and fundamental have comparable amplitude, resulting in what is usually referred to as a Type-B \textit{cathedral} QPO \cite{Casella2004}. Differently from the Type-C QPOs, which tend to be observable for several days or even weeks (albeit with an evolving centroid frequency), Type-B QPOs are transient features. Rapid transitions in which Type-B QPOs appear and disappear on sub-second time scales have been observed in a few sources (see for instance \cite{Nespoli2003, Casella2004} for the cases of GX 339-4 and XTE J1859+226). Type-B QPOs occur at a similar time to discrete jet ejections observed in radio \cite{Fender2004}, as radio flares and, sometimes, resolved jets in high angular resolution radio images \cite{Corbel2005}. This coincidence in time led to the suggestion that transient jets and type-B QPOs are causally connected. 
To date, the most accurate measurement of the time delay between the switch to a type-B QPO (observed with the \emph{Neutron Star Interior Composition Explorer}, {\it NICER}, \cite{Gendreau2016}) and the emergence of a radio flare (observed with the Arcminute Microkelvin Imager Large Array, AMI-LA, \cite{Zwart2008}) is of $\sim 2-2.5$ hr, as obtained from the multiwavelength monitoring campaign of the BH XRB MAXI J1820+070 \cite{Homan2020}. Observations at X-ray and radio wavelengths revealed the presence of relativistic ejections \cite{Bright2020,Espinasse2020,Wood2021} associated with these events. This represents the strongest empirical evidence to date that some physical connection may exist between the occurrence of a type-B QPO and the launch of jet ejecta. Such an inference was previously indirectly implied by the observed dependence of type-B QPOs strength on the viewing angle \cite{Motta2015}: this type of QPOs appear stronger in sources that are close to face-on, where also the observed jet power is expected to be stronger. However, \citet{Fender2009} showed that the association may not be straightforward, with the Type-B QPO sometimes occurring slightly before and sometimes slightly after the inferred ejection.
\\
\textbf{Type-A QPOs} are the least common type of LF QPOs in BH XRBs. The entire {\it RXTE} archive only contains $\sim 10$ significant Type-A QPO detections. These LFQPOs were initially dubbed type A-II by \cite{Homan2001}, following a classification proposed by \cite{Wijnands1999a}, where type A-I  QPOs were strong features associated with a low-amplitude red noise. Casella et al. \cite{Casella2005} later showed that type A-I QPOs should be classified as a Type-B QPOs.
Type-A QPOs normally appear in the soft state\footnote{Note that type-A QPOs have been considered part of the SIMS in some works. Here we follow the classification reported in \cite{Belloni2016,Ingram2019}.}, just after the SIMS to soft transition has taken place. They are observed as weak (few percent fractional rms) and broad peaked features ($Q \lsim 3$), with a centroid frequency $\sim~6-8$ Hz. Neither a sub-harmonic nor a second harmonic are usually detected, possibly because of the intrinsic low-amplitude of this type of QPO, or due to the broadness of the fundamental peak. Similarly to Type-B QPOs, Type-A QPOs are associated with a low amplitude broad band red noise component.

\subsection{High-frequency quasi periodic oscillations}

{\it RXTE}, with its unprecedented time resolution and the large collecting area, also made possible the discovery HF QPOs in BH systems (as well as of kHz QPOs in NS) \cite{vanderKlis1996,Morgan1997,Remillard1999}. Such very high frequency narrow features triggered great theoretical interest immediately after their discovery, most likely because  their centroid frequencies are comparable with the expected epicyclic frequencies of particle motion near the innermost stable circular orbit (ISCO) around a stellar mass BH or a NS  \cite{Abramowicz2001,Kato2004}. Although kHz QPOs from NS XRBs are common features \citep[e.g.,][]{Motta2017} with relatively high amplitudes, HF QPOs from BH XRBs are generally weak, and only little over ten detections exist in the entire {\it RXTE} archive \cite{Belloni2012}, if one excludes the HFQPOs from GRS 1915+105. 

The first HFQPO was  detected in 1997 in GRS 1915+105 (at $\sim 67$ Hz \cite{Morgan1997}) which is the source with the highest number of detections of HFQPOs, although it is not observed exactly at high frequencies, being generally found between 64 and 67Hz \cite{Belloni2013}. This QPO has also been detected by Astrosat many years after the RXTE detections at a very similar frequency \cite{Belloni2019}, indicating it represents a very fundamental frequency in this system. Other sources have been claimed to show HFQPOs - all at frequencies significantly higher than in GRS 1915+105 - are: XTE J1550$-$564, GRO J1655$-$40, XTE J1859+226, H 1743$-$322 , GX 339$-$4, XTE J1752$-$223, 4U 1630$-$47, IGR J17091$-$3624 (which also shows a high-frequency QPO at $\sim$67Hz, \cite{Altamirano2012}). However, most detections have been proven to be not statistically significant, leaving only a handful of significant detections from three sources (XTE J1550$-$564, GRO J1655-40, see \cite{Belloni2012}, and XTE J1859+226, Motta et al. submitted.).

Excluding the $\sim 67$ Hz, QPOs in GRS 1915+105 and IGR J17091$-$3624, all HFQPOs have frequencies of a few hundreds Hz, with typical quality factors $Q\sim 5-30$ and a relatively low amplitude, i.e. $F_{\rm var}\sim 0.5-6\%$ in the 2--40~keV band, increasing steeply with energy \cite{Belloni2013}. These features appear to be associated with high flux and intermediate states \cite{Belloni2012}, although this may be the result of a selection effect, as well as of an observing bias. HF QPOs can be observed as single or double peaks (in which case they are called the lower and upper HF QPOs), but only two BH XRBs - GRO J1655$-$40 and XTE J1859+226, showed two clear simultaneous peaks \cite[][and Motta et al. submitted]{Strohmayer2001,Motta2014}. The detection of two simultaneous HF QPOs in the light-curve from  XTE J1550$-$564 was reported, but later shown to be an effect of averaging a large number of observations \cite{Mendez2013}. The BH candidate H1743$-$322 also showed a significant HFQPO. A tentative simultaneous peak was also reported, which may correspond to a second HFQPO. Unfortunately, a firm classification was not possible owing to the low signal-to-noise ratio of the data \cite{Homan2005}.

\section{X-ray variability as a tracer of the accretion state}

The different accretion regimes are characterised by a drastic change in the nature of the dominant emission mechanism (e.g. thermal emission from the accretion disc, Comptonisation in an optically thin hot plasma). Therefore, spectral diagnostics are commonly used to identify and describe accretion states during an outburst \citep[e.g.][]{Fender2004,Homan2005b,Remillard2006}.
In this regard, the most widespread method relies on the representation in a HID (see Fig. \ref{fig:HID} and Introduction), where a source undergoing a complete outburst is observed to cycle through a well-defined ``q-shaped'' pattern, which is traversed always in the same anticlockwise direction. The position in the HID univocally corresponds to a given accretion regime. 

X-ray variability adds an important piece of information to the spectral picture, allowing for a more complete description of the different accretion states and a more precise detection of state transitions \citep[e.g.][]{Belloni2005a,Homan2005b}. 
Indeed, the different emitting regions contributing to the observed signal also display distinctive variability properties \citep[e.g.][]{Churazov2001,Gierlinski2005}. 
This can be intuitively deduced from the simple fact that the shape of the PDS depends on energy, with high-frequency (short time scale) variability noise components typically appearing or being more prominent at higher energies ($E\gsim$3 keV; e.g. \cite{Grinberg2014}). 
As a consequence, abrupt changes in the level of fast variability, as well as the appearance/disappearance of specific QPO types, characterise the evolution of the source throughout an outburst, as noticed since early studies of XRBs.

\begin{figure}
\centering
\includegraphics[width=0.45\textwidth]{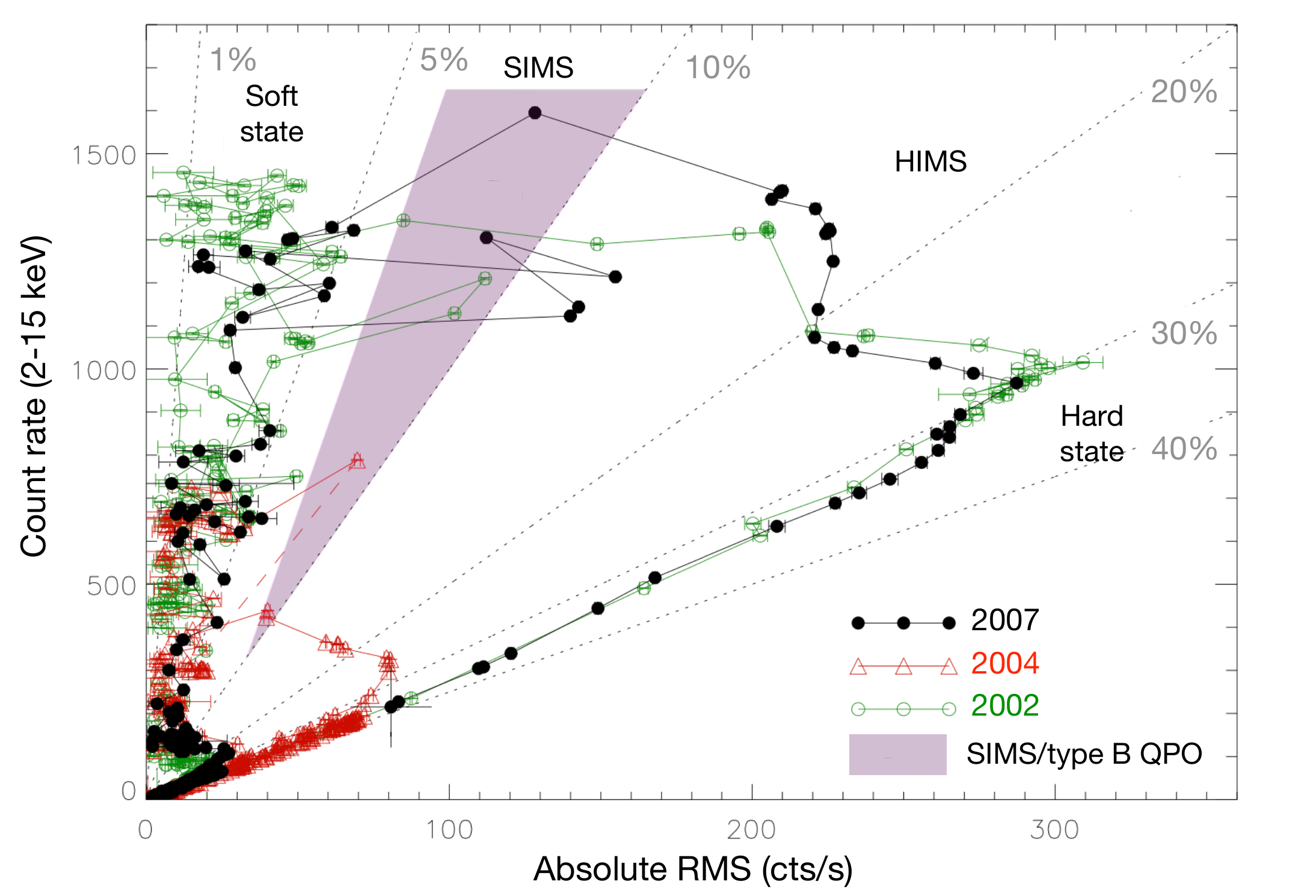}
\includegraphics[width=0.45\textwidth]{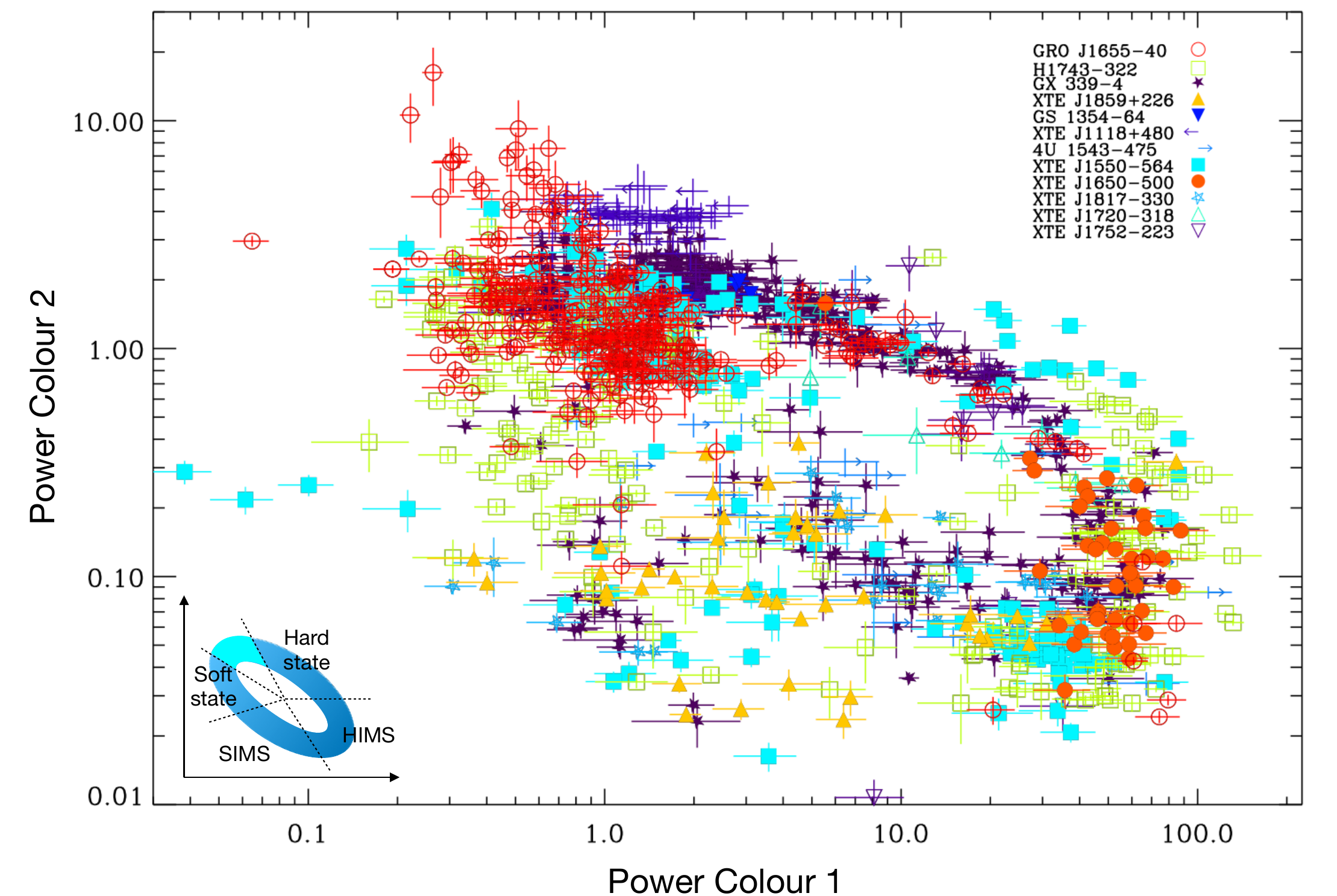}
\caption{Left panel: the RID of GX 339-4 as computed during three different outbursts. The different accretion states are marked in the plot according to the nomenclature adopted in this review. The percentages refer to the values of $F_{\rm var}$ measured in each state (in the frequency range $0.1–64$ Hz and in the $2-15$ keV energy range). Adapted from \cite{Munoz-Darias2011}. Right panel: the power colour-colour diagram (PC1 corresponds to the ratio of powers $P_{0.25-2.0 Hz}/P_{0.0039-0.031 Hz}$ and PC2 to $P_{0.031-0.25 Hz}/P_{2.0-16.0 Hz}$) of 12 BH XRBs observed by {\it RXTE}. The bottom-left graphics illustrates where the various states appear in the diagram (the cyan area represents an overlapping area between the hard and the soft states). Adapted from \cite{Heil2015}.}\label{fig:RID_PC}
\end{figure}

Similarly to the hardness parameter in HIDs, the rms variability amplitude can be used as a tracer of the different accretion regimes of the source by building a ``Rms-Intensity diagram'' (RID, Fig. \ref{fig:RID_PC}, left panel; \cite{Munoz-Darias2011,Heil2012}). In the RID the source is observed to follow different linear relations between the rms and the flux at different stages of the outburst, with the parameters of the relation depending on the accretion state of the source. 
An alternative model-independent approach for tracing accretion regimes through X-ray variability consists of comparing power-colours, namely ratios of total rms measured in different frequency bands (\cite{Heil2015}). Different sources consistently loop through a ``flattened doughnut'' in the power colour-colour diagram as they evolve throughout an outburst (Fig. \ref{fig:RID_PC} right panel). The state of the source can then be simply parametrised by the angle along the track, the so-called ``hue''.
Finally, the fractional rms, i.e. the fraction of flux contributing to variability in a given frequency range \cite{Belloni1990,Miyamoto1991a}, enables comparison between different sources and instruments, allowing us to quickly distinguish among hard, soft and intermediate states (Fig. \ref{fig:RID_PC}, left panel), although this parameter is insensitive to the exact phase of the outburst (e.g. similar values of fractional rms characterise the hard state both at the beginning and at the end of the outburst, being the luminosity dependence absent). In particular, hard states show high levels of fractional rms for energies $E\gsim$ 1 keV, i.e. $F_{\rm{var}}\sim20-50$ percent, distributed over temporal frequencies that range between $\sim$0.01--100 Hz. Prominent type-C QPOs are typically observed to be associated with such a strong broad band noise component. 
Soft states are instead characterised by remarkably low values of fractional rms, i.e. $F_{\rm{var}}\lsim$5 percent at $E\gsim$ 3 keV \cite{Munoz-Darias2011,Heil2012}. 
Intensive observational campaigns have shown that the $\lsim 10-20$ keV fractional rms gradually decreases \citep[e.g.][]{Gierlinski2005} between the low luminosity hard states at the beginning of an outburst and the high luminosity HIMS that precede the transition to the soft state (Fig. \ref{fig:RID_PC}, left panel). The opposite trend is observed at the end of an outburst, when the source leaves the soft state and returns to the hard state. 
Nonetheless, when the source transitions from the HIMS to the SIMS the fractional rms undergoes a sharp drop. At this transition, the broad band aperiodic X-ray variability becomes very weak (a few percent fractional rms) and any type-C QPO previously present in the PDS of the source is replaced by a single strongly peaked component, i.e. a type-B QPO \cite{Homan2020,Belloni2020}. 
These changes happen very quickly (down to seconds, see \cite{Nespoli2003}) thus they are rather difficult to catch (they can easily occur during gaps in the X-ray coverage). 
The best monitoring of a hard-to-soft state transition is to date the one obtained for the exceptionally bright (with peak luminosity of $\sim$4 Crab) transient system MAXI J1820+070 during its 2018 outburst. The monitoring (performed with {\it NICER}, \cite{Gendreau2016}) allowed the quick (a few hours) switch from a type-C to a type-B QPO to be witnessed for the very first time \cite{Homan2020}.

\subsection{A variable disc or a variable hard X-ray source?}
The significant drop of variability power observed between hard and soft states led to the conclusion that most of the rapid X-ray variability originates in the optically thin inner Comptonising region (which dominates the X-ray spectrum in the hard state), while the accretion disc (which is energetically dominant during the soft state), produces very little rapid variability \citep[e.g.][]{Churazov2001} in the frequency range typically investigated via Fourier analysis ($\sim 0.1-1000~Hz$). The idea of a highly unstable hard X-ray emitting region was corroborated by additional observational evidences. In particular, the fractional rms of the Comptonised tail at energies $E\gsim 10$ keV in the soft state can reach values as high or higher than those registered in the hard state \citep[e.g.][]{Gierlinski2005,Grinberg2014}. In addition, QPOs (and in particular the type-B QPO which represents the dominant variable component during the short-living SIMS) show high (low) fractional rms at hard (soft) energies, which are dominated by the Comptonised hard X-ray (thermal disc) emission \citep[e.g.][]{Sobolewska2006,Gao2014,Belloni2020}. This behaviour characterises also observations showing a prominent disc thermal component in the time-averaged spectra, strongly supporting the idea that the QPO mostly modulates the Comptonised hard X-ray emission.

Nonetheless, the idea of a negligible contribution from the disc to the X-ray variability properties of the source has been reconsidered with the availability of high throughput detectors with good timing capabilities, sensitive at $E<1$ keV (i.e. the EPIC-pn onboard XMM-Newton, and {\it NICER}'s X-ray Timing Instrument), which allowed X-ray timing studies of the disc in the hard state to be performed.
In the hard state the disc is cooler ($T_{\rm{in}}\sim0.2-0.3$ keV) and dimmer, giving its most significant contribution to the softest part of the X-ray spectrum.  
Unexpectedly, these studies revealed high levels of intrinsic disc variability on time scales longer than $\sim$ 1-10 s. 
On these relatively long time scales the disc ($E\lsim$1 keV) can be more variable than the $\sim$2-10 keV Comptonisation component \cite{Wilkinson2009,DeMarco2015a} during most of the hard state. This demonstrates that the variability properties of the optically thick and geometrically thin disc change significantly between the hard and the soft state. In other words the disc is likely to be intrinsically unstable in the hard state, which invites speculations on the role played by such a difference in the properties of the disc, and in the formation/quenching of a strong hard X-ray source in hard/soft states.
Conversely, on short time scales ($\lsim1$s) the disc in the hard state typically varies less than the power law emission. This is in line with the rapid disc variability being triggered by variable X-ray heating. 
A similar long/short time scale behaviour has been observed in comparative X-ray/optical studies of some AGN \citep[e.g.][]{Arevalo2009,Alston2019}.

Interestingly, while not dominant, contribution from variable disc emission has been also found in association with QPO modulations. In particular, phase-resolved spectroscopic studies of both type-C and type-B QPOs in a few sources (e.g. H1743-322 and GX 339-4) revealed signatures of disc-reprocessing of the quasi-periodically variable hard X-ray signal, in the form of a rocking (from blue to red shifted) Fe K-$\alpha$ line \cite{Ingram2016,Ingram2017,Nathan2022} and of modulated variations of the disc black body temperature \cite{Stevens2016}. These findings support models that invoke a geometric origin for QPOs (e.g. Lense-Thirring precession of the inner hot flow; \cite{Stella1998,Ingram2009,Fragile2007}).

\section{X-ray cross-spectral-timing studies of BH XRBs}

Understanding the causal connection between the different emitting regions is key in order to unveil where variability is produced and how it can be transferred from one region to the others. An irradiated gas will respond to changes of the primary source in a way that can be described in terms of the \emph{impulse response function}\footnote{The term \emph{transfer function} is more typically used in the Fourier-frequency domain.} associated with that particular system and radiative process. The impulse response function theoretically describes how a pulse of primary photons is redistributed by the irradiated gas over time and energy. The response function depends on the physical parameters of the gas. Therefore, theoretical impulse response functions allow us to infer physical constraints from the timing and spectral information extracted from the data \citep[e.g.][]{Poutanen2002}.

Constraining the response function that best describes the physical system under study is among the main goals of cross-spectral timing analysis.
Cross-spectral techniques are based on the computation of the Fourier cross-spectrum. Given light curves $x(t)$ and $y(t)$ extracted in two different energy bands, the cross-spectrum is defined by
 $$
 C(\nu)=X^{\ast}(\nu)Y(\nu),
 $$
 where $X(\nu)$ and $Y(\nu)$ are the Fourier transforms of the two light curves, and the asterisk denotes the complex conjugate (e.g. \cite{Uttley2014} and the chapter "Basics of Fourier Analysis for High-Energy Astronomy" in this book).
The cross-spectrum is a complex quantity, thus it can be represented as a vector in the complex plane. Given the stochastic nature of variability in BH-accreting systems, a consistent estimate of the intrinsic cross-spectrum can be obtained only by averaging over different realisation (i.e. sample light curves) of the underlying process. This can be done by extracting multiple pairs of light curves from consecutive data segments, measuring the cross-spectrum for each pair, and averaging over the different cross-spectra. In the complex plane \cite{Nowak1999} this is equivalent to summing vectors in order to infer a dominant direction  - which defines the phase difference between the two light curves and the corresponding time lag - and the amplitude of the resulting vector - which yields their degree of coherence, i.e. the fraction of linearly correlated variability. Coherence and time lags between X-ray bands in BH XRBs will be discussed in the following sections (see also the chapter "Basics of Fourier Analysis for High-Energy Astronomy" in this book). If the light curves are chosen so as to be each dominated by emission from different parts of the accretion flow, these measurements allow us to study the causal relationship between these physical regions.

The Fourier cross-spectrum is the frequency domain-analogue of the cross-correlation function (``cross-correlation theorem'', see the chapter "Basics of Fourier Analysis for High-Energy Astronomy" in this book), but it has the advantage to depend on the temporal frequency, allowing  potentially complex time-dependent behaviours to be easily unveiled. Thus, the Fourier cross-spectrum is particularly suitable to study physical processes that depend on a time scale. 
Moreover, given that faster (slower) variability originates in the inner/smaller (outer/more extended) regions \cite{Churazov2001}, this tool allows us to analyse the multi-scale structure of the accretion flow by singling out emission produced at different distances from the BH.

\subsection{Coherence}

The coherence is a powerful albeit often overlooked tool to discriminate among physical models of variable X-ray emission.
The coherence quantifies the degree of linear correlation between two light curves. 
In more technical terms (see also \cite{Vaughan1997,Uttley2014} and the chapter "Basics of Fourier Analysis for High-Energy Astronomy" in this book), the coherence function is defined by 
$$
\gamma^2=\frac{|\langle C(\nu) \rangle|^2}{\langle |X(\nu)|^2 \rangle \langle |Y(\nu)|^2 \rangle},
$$

where $C(\nu)$ is the cross-spectrum (see previous section), and the denominator is the product of the power spectra of the two light curves. The angle brackets in the equation denote average values. Note that $\gamma^2$ is a statistical quantity, thus defined over a number (possibly large) of different realisations of the variability process. If the two processes are fully coherent ($\gamma^2=1$) a linear transformation exists between them (i.e. a linear response function), with the transformation being the same for all the realisations of the process (i.e., sample light curves). For a detailed discussion of the statistical properties of the coherence function we refer to \cite{Bendat1986,Vaughan1997,Nowak1999}.
An intuitive description of the coherence derives from the complex plane representation of cross-spectrum vectors discussed in the previous section. If two light curves are fully coherent, the corresponding cross-spectrum vectors will result perfectly aligned. The presence of an incoherent variability component (due to Poisson noise or to non-linear physical processes) adds up as a vector in the complex plane with a phase randomly and uniformly distributed over $[-\pi:\pi]$. This contribution will introduce a scatter in the direction of the cross-spectrum vectors of the intrinsically coherent processes, thus decreasing the resulting coherence ($\gamma^2<1$)\footnote{Note that for incoherent processes the true coherence is zero, but the measured coherence depends on the number of averaged realisations. If this number tends to $\infty$ than $\gamma^2\rightarrow0$, but for a finite number of realisations, in general $\gamma^2$ will be small but $>0$.}. The coherence of noisy data will always be less than unity, but in many frequency regimes the effect of counting noise can be accounted for and the intrinsic coherence easily inferred \cite{Vaughan1997}.

A large number of physical processes can lead to a loss of coherence. For example reflection of hard X-ray photons in the accretion disc gives rise to a spectrum which includes contribution from electron scattering as well as absorption and emission lines \cite{Guilbert1988}, each process dominating at different energies. While the Compton scattering component in the reflection spectrum responds linearly to variations of the irradiating flux, the response of absorption/emission components is non-linear due to the induced changes in the ionisation level of the irradiated disc atmosphere \citep[e.g.][]{DeMarco2020,Mastroserio2021,Juranova2022}. 
Other examples of non-linear processes include changes of the disc black body temperature in response to irradiation of a variable primary continuum \cite{Vaughan1997}, spectral pivoting of the hard X-ray power law \cite{Mastroserio2018}, changes in size or temperature of the Comptonising region (on time scales longer than the duration of the single realisations of the process; \cite{Nowak1996}), uncorrelated flares from multiple discrete regions of the accretion flow \cite{Vaughan1997}. 
Nonetheless, studies of BH XRBs generally show high levels of coherence across X-ray energy bands and over a large range of time scales (roughly between $\nu\sim 10^{-2}-10$ Hz) \citep[e.g.][]{Nowak1999,Wilkinson2009,Grinberg2014}. 
High coherence is generally observed in the Comptonisation-dominated parts of the spectrum \citep[e.g.][]{Grinberg2014}.
For example, in Cyg X-1 a coherence level consistent with unity is measured during hard and soft states between energy bands $\gsim$2 keV, while this value slightly decreases during intermediate states, e.g. \cite{Cui1997,Grinberg2014}.
This suggests that the dominant processes at work are linear or that any contribution from non-linear processes can be well approximated at first order by a linear response function (this applies when the perturbations in the parameters contributing to non-linear terms are relatively small; \citep[e.g.][]{Kotov2001,Koerding2004,Mastroserio2018,Mastroserio2021}).

High levels of coherence are also observed between the Comptonisation component and the variable fraction of thermal disc emission \cite{Wilkinson2009}\footnote{In this paper the covariance spectrum is measured instead of the coherence function. The two are closely related, in that the covariance spectrum gives the spectral shape of linearly correlated components, \cite{Uttley2014}.}, at least during the hard state. 
As previously discussed, on time scales $\gsim$ 1s, the disc is more variable than the power law during most of the hard state \cite{Wilkinson2009,DeMarco2015a}, suggesting that the slower variability is intrinsically produced in the disc, and it may be (partially) transferred to the X-ray source in such a way that a high degree of coherence is preserved. 
On the other hand, the disc is less variable than the power law on time scales $\lsim1$s during the early phases of the outburst (and even on longer time scales in the bright hard state, \cite{DeMarco2015a}). The high observed coherence on these time scales therefore may be explained if the most rapid variations in the disc are triggered by X-ray heating, whereby the variable hard X-ray photons thermalise in the disc producing variable thermal emission (changes in the normalisation of the thermal component preserve coherence \cite{Vaughan1997}). 

In general, the coherence appears to decrease at very high temporal frequencies (e.g. \cite{Grinberg2014}), which might indicate the presence of multiple, causally disconnected flaring regions \cite{Nowak1999}. However, at these frequencies the effects of counting noise become dominant, thus the intrinsic value of coherence (free from counting noise effects) is generally not accurately determined \cite{Vaughan1997}.
As discussed in the next section, X-ray lags can give additional important clues on the physical origin of the observed high coherence.

\subsection{X-ray time lags}

Phase lags between two light curves can be computed as the argument of the cross-spectrum $\phi(\nu)=arg[C(\nu)]$ (units of radians). The phase lag is related to the time lag $\tau$ (units of seconds) by $\tau(\nu)=\phi(\nu)/2\pi\nu$ (see also the chapter "Basics of Fourier Analysis for High-Energy Astronomy" in this book). 
As previously explained, in the complex plane representation, the direction of the resulting average cross-spectrum vector defines the phase lag $\phi$ between two light curves. 

Time lags are naturally expected in complex systems characterised by strong variability. For example, time lags associated with diffusion/scattering processes or light travel time differences can arise as a consequence of the extent/physical separation of the emitting regions in the accretion flow \cite[e.g.][]{Poutanen2002}. Time lags can also be due to the time needed for a gas to physically readjust to changes of the irradiating flux (for example via photoionisation and recombination; e.g. \citep[e.g.][]{Nicastro1999,Silva2016,Juranova2022}). 

X-ray lags in BH XRBs have been extensively studied, particularly in the hard state where more sensitive measurements  are possible thanks to the combined high fractional rms and high count rates registered in these states (\cite{Uttley2014}). These studies revealed the presence of X-ray lags which strongly depend on the energy and temporal frequency. 
X-ray lags associated with both aperiodic variability components and QPOs are commonly observed, though the physical nature of these lags might differ. 
At low frequencies ($\lsim2$ Hz), the aperiodic variability in softer bands is observed to lead the aperiodic variability in harder bands, producing ``hard lags'' (e.g. \cite{Nowak1999,Grinberg2014,Uttley2011,DeMarco2015}, Fig. \ref{fig:MAXILags}, left panel). 
At high frequencies ($\gsim2$ Hz),
``X-ray reverberation lags'' due to the time delayed response of the disc to irradiation from rapidly variable hard X-ray photons are being detected in an increasingly larger number of sources \cite{Uttley2011,DeMarco2015,DeMarco2016,DeMarco2017,Kara2019,DeMarco2021,Wang2021,Wang2022} (Fig. \ref{fig:MAXILags}, left panel). 
Lags associated with QPOs are also commonly detected. In this case the lag is measured in a small frequency range centered around the QPO centroid frequency. QPO lags show peculiar switches in sign (i.e. changing between hard photons lagging soft photons, and soft photons lagging hard photons) as a function of QPO frequency and inclination of the source \citep[e.g.][]{Reig2000,Gao2014,vandenEijnden2017}. In the following sections we will discuss the phenomenology of these different classes of X-ray lags and some of the proposed interpretations.  

\subsubsection{Hard X-ray lags of the aperiodic variability}

Hard X-ray lags are commonly observed in BH XRBs during the hard state and the HIMS. In Cyg X-1 measurements of hard lags have been also reported in the soft state \cite[e.g.][]{Pottschmidt2000,Grinberg2014}, although the soft state typically observed in Cyg X-1 does not completely coincide with the soft state in more ordinary BH XRBs.

Due to sensitivity limitations of available instruments, early studies of the hard X-ray lags in BH XRBs, the great majority being performed with {\it RXTE},  were  restricted to energy bands dominated by the Comptonised emission ($E\gsim 1$ keV; e.g. \cite{Miyamoto1988,Nowak1999}). 
The amplitude of these hard lags is observed to strongly depend on the temporal frequency of variability, following a $\tau \propto \nu^{-0.7}$ trend, although not smoothly, as humps and steps clearly appear in high quality data \citep[e.g.][]{Nowak1999,Uttley2011,DeMarco2015,DeMarco2021}. Very long hard lags of 1-10 s have been measured at the lowest sampled frequencies, while at the highest frequencies the lag amplitude drops to values of the order of milliseconds or smaller \citep[e.g.][]{Miyamoto1988,MunozDarias2011a,DeMarco2015}. 
Hard lags also display an approximately log-linear dependence on energy, with longer amplitudes measured between more distant energy bands \citep[e.g.][]{Nowak1999,Uttley2011,DeMarco2015}. 

Detailed analyses demonstrated that hard lags in BH XRBs are intrinsic to the primary hard X-ray continuum. In particular, these studies excluded that the hard X-ray reflection components from the disc (i.e. the FeK line and the Compton hump) could give any dominant contribution to the low-frequency lag spectrum \cite{Kotov2001,Cassatella2012}, although some minor features due to disc reprocessing can still appear, superimposed on the low-frequency continuum hard lags \cite{Mastroserio2018}. 

Inverse Compton up-scattering, responsible for the bulk of the hard X-ray emission in BH systems, was instead discussed as a possible production mechanism for the hard X-ray lags, due to the fact that this process naturally produces hard lags (as harder photons undergo a larger number of scatterings than soft photons in order to reach the observed energies).
However, Compton scattering in an electron cloud of uniform density and temperature does not produce lags that depend on Fourier frequency, a fact that is at odds with observations  \citep[e.g.][]{Miyamoto1988,Kazanas1997}. In addition, the large range of observed lag amplitudes requires the scattering region to extend up to hundreds of gravitational radii ($R_g$) in order to explain the longest ($\sim$1-10 s) observed lags (as can be inferred for example using equation 20 in \cite{Nowak1999}). Furthermore, \cite{Maccarone2000} noticed that variable Comptonisation in an extended corona would produce an auto-correlation function with a width that increases with increasing energy, again at odds with observations which show that the slower variations are produced at lower energies.  
These limitations can be overcome if more complex Comptonisation models are taken into account. For example, it has been proposed that Comptonisation within a mildly relativistic jet (as seen in the hard state of BH XRBs), with electron density in the jet inversely proportional to the distance, can explain the observed hard lags \cite{Giannios2004}. By letting the radius of the base of the jet and the Thomson optical depth along the jet axis vary, Comptonisation in the jet can explain the observed \cite{Grinberg2014,Altamirano2015} correlation between the hard lag amplitude and the photon index of the power law component in the spectrum \cite{Reig2018,Kylafis2018}, as well as the inclination-dependent scattering of this correlation (the correlation is tighter in low-inclination systems; \cite{Reig2019}). 
However, according to the model, the size of the jet base, which is smaller than the transition radius between the hot inner flow that feeds the jet and the optically thick outer accretion disc, should reach quite large values (of the order of 300 $R_g$) at the transition to the soft state \cite{Reig2019}. 

An alternative explanation is that the hard lags are intrinsic to the diffusion mechanism in the accretion flow. Such a possibility was proposed as soon as the first detections of hard X-ray lags were reported \cite{Miyamoto1988,Miller1995,Nowak1996}. This is the core idea of the class of models referred to as ``propagating mass accretion rate fluctuations'' models \cite{Lyubarskii1997,Kotov2001,Arevalo2006,Ingram2013}. These models assume that fluctuations in the mass accretion rate occur independently at different radii, and that a fraction of these local fluctuations propagates downstream, modulating fluctuations produced in the inner radii. The fact that diffusion time scales can be quite long at large radii implies that at these scales the most rapid changes are quickly suppressed, and only longer-lived fluctuations can propagate down from these distant regions. When these modulations reach the main inner zones of energy release, they are emitted as variable X-ray flux. The resulting variability is thus a consequence of the combined contribution of modulated fluctuations produced throughout the flow and characterised by widely different time scales of variability (in agreement with the large dynamical range of time scales observed in the power spectrum of these sources).
For the propagating mass accretion rate fluctuations model to produce hard X-ray lags, the innermost emitting regions should be spectrally stratified (i.e., the radial emissivity profile should be energy-dependent), with more hard X-ray photons produced close to the BH, and more soft X-ray photons produced farther out \cite{Kotov2001,Mahmoud2018}. A mass accretion rate perturbation propagating inwards would then naturally imprint hard lags in the observed variable flux. The amplitude of the lag depends on the diffusion time scale, the emissivity profile, and the physical separation between the emitting regions. Fluctuations produced within the emitting region are responsible for the observed frequency-dependence of hard lag amplitudes, with the longest lags due to propagation from more distant regions, and thus associated with slower variability components. As a consequence, the model requires a radially extended X-ray emitting region/hot flow in order to resemble the data \cite{Arevalo2006,Rapisarda2017,Mahmoud2018,Mahmoud2019}.
However, while reproducing the overall shape of the lags, fits with available models still show quantitative discrepancies with the data \citep[e.g.][]{Rapisarda2017a,Mahmoud2018}, possibly calling for some modification of the paradigm \cite{Veledina2018}.

It is worth noting that a variable flux of thermal seed photons irradiating a hot plasma cloud will change the plasma temperature and the slope of the resulting X-ray power law. Therefore, intrinsically variable thermal disc emission is expected to trigger spectral variability of the primary hard X-ray continuum. It has been shown that these spectral variations can potentially contribute to the observed hard lags \citep[e.g.][]{Poutanen1999,Koerding2004,Uttley2014,Mastroserio2018}.

\begin{figure}
\centering
\includegraphics[width=0.48\textwidth]{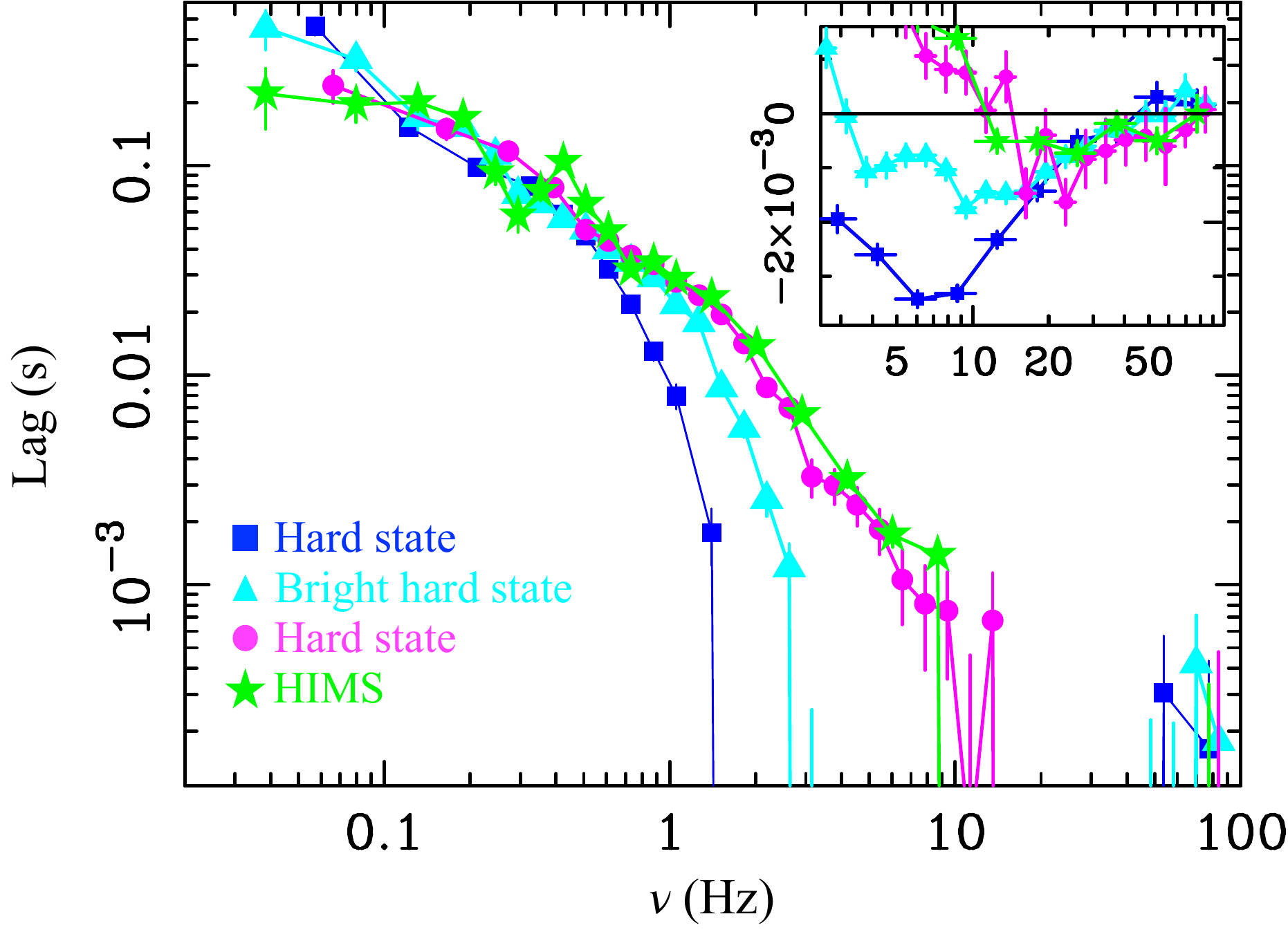}
\includegraphics[width=0.47\textwidth]{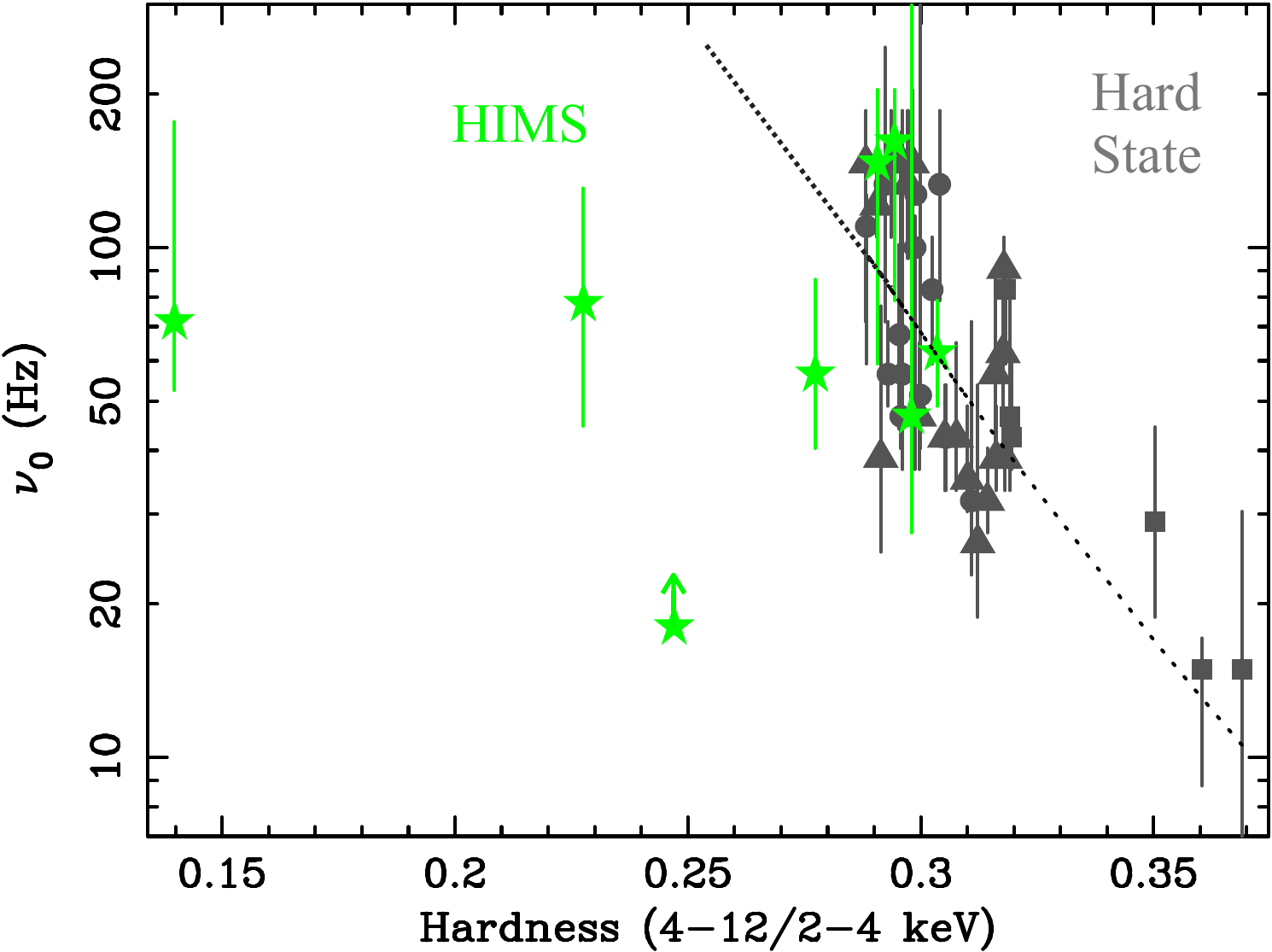}
\caption{Left panel: The 0.5–1 keV (disc-dominated) vs. 2–5 keV (power law-dominated) lag-frequency spectrum of selected {\it NICER} observations of MAXI J1820+070 during different phases of its 2018 outburst, showing the low-frequency hard lag and the high-frequency reverberation lag (inset). Right panel: The evolution of the X-ray reverberation lag's characteristic frequency in MAXI J1820+070. This trend is indicative of decreasing relative distances between the X-ray source and the disc in the hard state, and a sudden increase at the beginning of the HIMS. Adapted from \cite{DeMarco2021}}\label{fig:MAXILags}
\end{figure}

In more recent years, thanks to the availability of high throughput instruments sensitive to energies down to $\sim$0.3 keV (e.g., the EPIC-pn instrument on board XMM-Newton), it has been possible to extend the analysis of X-ray lags to energies where the disc can be directly observed even in the power law-dominated hard state \citep[e.g.][]{Tomsick2008}, and its causal relationship with the other spectral components directly studied (Fig. \ref{fig:MAXILags}).
These studies have shown that the log-linear dependence of hard X-ray lags on energy becomes significantly steeper in the disc-dominated soft band. This steepening is observed in the slower variability components (time scales $\gsim$ 1s), and indicates that the slow variability of the Comptonised hard X-ray flux lags behind the slow variability of the disc with a delay longer than that measured among power law-dominated bands with similar energy separations \cite{Uttley2011,DeMarco2015}. 
The amplitude of these disc-to-power law lags show the same inverse scaling with frequency as that characterising the hard lags intrinsic to the power law, at least as long as temporal frequencies $\nu\lsim0.1-1$ Hz are considered \cite{Uttley2011,DeMarco2015}, suggesting a similar underlying physical mechanism.
In the context of propagating fluctuations models, the relatively longer disc-to-power law lags might be due to a diffusion time from the disc to the Comptonising region longer than within the Comptonising region itself. Long time duration General relativistic magnetohydrodynamical (GRMHD) simulations of optically thick and geometrically thin flows show that fluctuations of mass accretion rate can diffuse down in standard accretion discs, driving radially-coherent changes of mass accretion rate \cite{Hogg2016}. These fluctuations may thus modulate variability produced within the hot, optically thin and hard X-ray emitting inner flow, thus explaining the broad energy range (including disc-dominated energy bands) where hard lags can be observed.

\subsubsection{X-ray reverberation lags}

The frequency-dependence of hard X-ray lags implies that at high frequencies ($>$1 Hz) their contribution becomes almost negligible. At these frequencies, when time lags are measured with respect to a disc-dominated energy band, the emergence of a short (i.e. with amplitude of tens-to-a few milliseconds), soft X-ray lag is observed in a number of BH XRBs in the hard state in high signal-to-noise data \citep[e.g.][]{Uttley2011,DeMarco2015,DeMarco2016,DeMarco2017,Kara2019} (Fig. \ref{fig:MAXILags}, left panel).  
These lags have been interpreted as due to the thermal response of the disc after rapid variations of the primary hard X-ray irradiating flux. In the literature they are referred to as "X-ray reverberation lags" (or sometimes ``X-ray thermal reverberation lags'' to highlight the fact that they are associated with the thermal response of the disc), and are predicted as a natural consequence of corona-disc feedback processes. 
Given the high gas densities, reprocessing of hard X-ray photons should occur almost instantaneously in the disc, therefore it is assumed that the amplitude of the reverberation lag is mostly determined by the time needed for the hard photons to travel the additional light path that separates the hard X-ray source from the reprocessing region in the disc. As such, they can be converted into absolute distances, thus providing a direct diagnostic of the geometry of the hard X-ray source and the inner accretion disc \cite{Uttley2014}.

The first indication of thermal X-ray reverberation in a BH XRB came from the study of two consecutive XMM-Newton observations of GX 339-4 during its hard state \cite{Uttley2011}. 
This interpretation was supported by the analysis of covariance spectra, which showed that rapid disc variability in the hard state is linearly correlated with variable hard X-ray photons \cite{Wilkinson2009} as expected if a causal connection exists between the two. Subsequent studies digged in the XMM-Newton archive, revealing the presence of a thermal reverberation lag in seven more observations of GX 339-4 and in four observations of H1743-322 \cite{DeMarco2015,DeMarco2016,DeMarco2017}, one of these also showing hints of reverberation in the Fe K-$\alpha$ line component of the reflection spectrum \cite{DeMarco2017}.
The advent of {\it NICER} significantly increased the number of detections of X-ray reverberation lags in BH XRBs \cite{Kara2019,Wang2020,DeMarco2021,Wang2021,Wang2022}, thanks to its higher effective area at soft energies, good timing capabilities, and its ability to observe very bright sources with negligible instrumental effects.  

The systematic study of X-ray reverberation in multiple observations of the same source turned out to be also crucial to unveil trends of lag amplitude as a function of the accretion state (Fig. \ref{fig:MAXILags}). Changes in the geometry of the innermost accretion flow are thought to play a major role in the observed evolution of BH XRBs throughout an outburst \citep[e.g.][]{Esin1997}, thus the observed trends have been interpreted as a natural consequence of such changes.
In particular, in GX 339-4, the X-ray reverberation lag is observed to become shorter as the luminosity of the source increases during the hard state \cite{DeMarco2015,DeMarco2016,DeMarco2017}. This suggests that the relative distance between the X-ray source and the reprocessing region in the disc decreases as the source approaches the peak of luminosity in the hard state. 
The higher quality {\it NICER} data, allowed the X-ray reverberation lag to be tracked in a much larger number of observations of the source MAXI J1820+070, covering the entire hard state and the intermediate states (HIMS) preceding the transition to the soft state \cite{DeMarco2021} (Fig. \ref{fig:MAXILags}, right panel). In the hard state the lag's intrinsic amplitude systematically decreases (down to values of a few milliseconds) as a function of spectral hardness \cite{Kara2019,DeMarco2021}, implying a gradual decrease of relative distances between the X-ray source and the disc as the outburst develops, in agreement with the results from GX 339-4.
Nonetheless, the lag suddenly becomes longer (reaching values $\gsim 0.02$s) about 4 days before the transition to the soft state \cite{DeMarco2021,Wang2021} (Fig. \ref{fig:MAXILags}, right panel). This event precedes the emergence of a strong radio flare and the ejection of relativistic plasma \cite{Homan2020,Bright2020,Espinasse2020,Wood2021}, suggesting a possible causal connection between the two phenomena. Qualitatively these results can be explained assuming that right before the hard-to-soft transition dissipation of hard X-rays predominantly occurs in a larger or more distant region, possibly associated with the expelled ballistic jet. 

Observations of decreasing lag amplitude in the hard state, on the other hand, are in line with expectations of truncated disc/hot inner flow models. Indeed, the observed trends suggest that the disc inner radius gradually moves inwards between the high disc-truncation configuration characterising the low luminosity hard state \citep[e.g.][]{Esin1997,Meyer1994} and the standard disc configuration (inner disc radius close to/at the ISCO) characterising the soft state \cite{Shakura1973}.
Nonetheless, an alternative scenario has been proposed, whereby the observed trend of lag amplitude is mostly due to changes of the geometry the X-ray source, with the hot Comptonising plasma contracting above a disc, which, differently from the previous scenario, extends close to the ISCO during most of the hard and intermediate states \cite{Kara2019,Wang2022}.

Studying X-ray reverberation lags in an increasingly higher number of BH XRBs and accretion states is of the utmost importance in order to better understand how these systems evolve throughout an outburst. However, the development of X-ray spectral-timing models is also crucial in order to obtain strong constraints on the inner truncation radius of the disc and the geometry of the X-ray source. A notable effort has been recently put in the development of models able to simultaneously fit the X-ray reverberation lags and the hard lags intrinsic to the broad band continuum. In most cases these models are designed in order to fit also other spectral-timing products (e.g. the rms spectrum) and/or the time-averaged spectrum, thus reducing the degeneracies in the models. 
The main differences among the proposed models are in the prescription for the mechanism responsible for the production of hard X-ray lags. The majority of these models assume hard X-ray lags to be due to propagating fluctuations in an extended hot flow (e.g. PROPFLUC, \cite{Ingram2011,Ingram2012,Rapisarda2014}). However, other possibilities have been also considered, such as a pivoting power-law continuum (RELTRANS, \cite{Mastroserio2018,Ingram2019a,Mastroserio2021}) and Comptonisation in an extended X-ray source, the latter being developed to explain the lags associated with QPOs \cite{Karpouzas2020,Garcia2022} (see next section), but possibly extendable to partly explain the hard lags associated with the broad band noise components.

\subsubsection{Lags associated with QPOs}

A number of properties of QPOs - like their fractional amplitude and
 quality factor - tend to correlate with the QPO frequency and with the accretion state of the source. Hence, much of the work done on QPOs over the past two decades has focused on the characterisation of the relation between characteristic frequencies and the dependence of such frequencies on other source properties. 
In this context, an aspect of the X-ray signal that has received renewed attention in the recent years is the energy- and frequency-dependent time (or phase) lags associated with a QPO (i.e. calculated over a small frequency range centered around the QPO frequency \citep[eg][]{Casella2005}).

\begin{figure}
\centering
\includegraphics[width=1.0\textwidth]{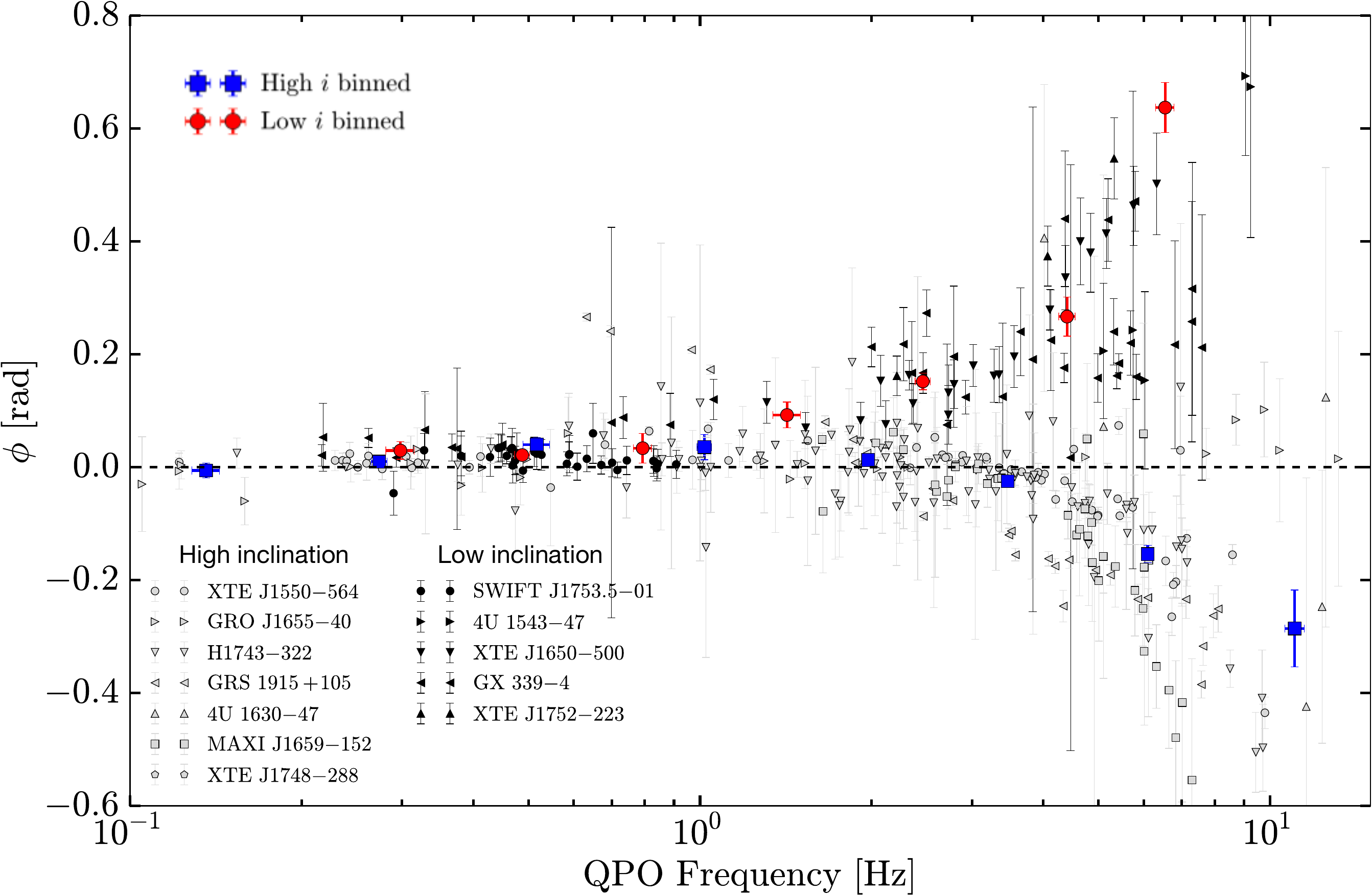}
\caption{Type-C QPO phase lags as a function of the QPO frequency. Black and grey points indicate low- and high-inclination sources, respectively. The red and blue points show the average phase lag and the QPO frequency in logarithmic frequency bins. Adapted from \cite{vandenEijnden2017}. } \label{fig:QPO_lags}
\end{figure}

The different types of low-frequency QPOs have been known to show different phase lags at the QPO fundamental and (sub)harmonic \citep[e.g][]{Pahari2013}, and the relation between the QPO frequency and the QPO phase lag differs among sources \citep[see, e.g.,][]{Reig2000}). A particularly noteworthy example is the apparent log-linear dependence of QPO phase lags on the QPO frequency observed in  GRS 1915+105 \cite{Reig2000, Qu2010, Pahari2013}. 
\cite{Motta2015} showed that the rms amplitude of the type-C QPO as a function of QPO
frequency is higher in sources observed at high inclination angles to the line of sight (i.e. more edge-on), while the opposite is true for type-B QPOs\footnote{The number of known Type-A QPOs is too small to allow a systematic study of their population properties to be performed.}. \cite{vandenEijnden2017} studied the same sample of sources considered by \cite{Motta2015} plus GRS 1915+105, and showed that in high inclination sources, above a certain QPO frequency, the phase lags measured at the type-C QPO frequency are negative (soft lags, i.e. the low-energy photons lag the high-energy photons), while in low inclination source the phase lags of the type-C QPOs are positive (hard lags). In both cases the lag amplitude increases in its absolute value with the QPO frequency (see Fig. \ref{fig:QPO_lags}). These authors also found that the type-B QPO soft lags are associated with high-inclination sources, although the available source sample is still too small and such a result needs to be confirmed by additional data. These results support a geometrical origin for the type-C QPOs, as previously suggested by other authors \citep[e.g.][]{Schnittman2006, Ingram2009, Motta2015} and point to a different origin for type-B and type-C QPOs. Furthermore, such findings are consistent with what is expected in case the phase lags originate from a pivoting energy spectrum which changes over the QPO cycle, while the inclination dependence arises from differences in the dominant relativistic effects.

In particular, the phase lags associated with the type-C QPOs in the source GRS 1915+105 have been extensively studied in the literature. \cite{Pahari2013, Qu2010, Reig2000} found that the type-C QPO phase lag smoothly varies with energy, something that has been observed in the BH binary XTE J1550-564 as well \cite{Wijnands1999}. In the case of GRS 1915+105, \cite{Qu2010} found that the slope of the lag energy dependence changes from positive when the QPO frequency is found below $\approx$2 Hz (i.e. hard photons lag soft
photons) to negative when the QPO frequency moves above that frequency \citep[see also][]{Qu2010,Pahari2013}. Based on the fact that the type-C QPO frequency in GRS 1915+105 had previously been observed to depend on photon energy, \cite{vandenEijnden2016} studied further the energy-dependence properties of this QPO,  and found that the phase lag between two broad energy bands generally increases for a number of QPO cycles, then the QPO becomes decoherent, the phase lag becomes zero, and the pattern repeats. The faster the QPO becomes decoherent, the faster the phase lag increases. The above suggests that the oscillation in the harder band is faster than (``runs away from'') the softer band oscillation, and that the resulting frequency difference contributes to the decoherence of the QPO. The above indicates that the energy dependence of the QPO frequency is indeed an intrinsic property of the QPO itself.


While the literature includes several works devoted to the study of lags associated with low-frequency QPOs, only a few works attempt to explain quantitatively their energy or frequency dependence in the context of radiative mechanisms. \cite{Nobili2000} proposed a Comptonization model consisting of a non-isothermal Compton cloud filling the inner parts of the accretion flow (and becoming denser as the truncated inner disc radius moves inwards). The model reproduces the observed switch from hard to soft lags as a consequence of Compton up- and down-scattering, respectively becoming the dominant mechanism as the density of the inner and hotter parts of the Compton cloud increases.

\cite{Karpouzas2020} proposed a model to explain the soft lags associated with lower kHz QPO in  NS low mass XRBs in terms of a delayed heating of the seed photons by photons previously up-scattered in the corona. Using the same model, \cite{Garcia2021} showed that a two-component extended corona explains the energy dependence of the time lags and rms amplitude of the type-B QPO in MAXI J1348-630.  \cite{Karpouzas2021} extended the analysis by \cite{Garcia2021} to include type-C QPOs. They considered the type-C QPOs observed in GRS 1915+105 and found that the phase lags of the QPO make a transition from hard to soft at around 1.8 Hz, as already observed by \cite{vandenEijnden2016} and \cite{vandenEijnden2017}. This is consistent with a large corona of $\approx$ 100~R$_g$ that covers most of the accretion disc when the QPO frequency is large. As the observed QPO frequency decreases, the corona shrinks down to $\approx$ 10~R$_g$, until the heating of the disc by the corona becomes inefficient when the QPO frequency moves below 1.8 Hz, resulting in to hard lags.

\section{A brief comparison between BH XRBs and AGN variability}

The X-ray variability of (both BH and NS) XRBs and AGN shows notable similarities, but also puzzling differences.
AGN light curves display the same kind of stochastic variability as seen in BH XRBs. Due to their large BH masses, a typical continuous X-ray observation only samples the highest frequency part of the PDS of an AGN. In this regime the PDS of AGN can be simply modelled by a broken (or bended) power law \citep[e.g.][]{Uttley2002,Gonzalez2012}. The break (bend) frequency of such a power law is observed to scale inversely as a function of BH mass and directly as a function of the accretion rate \cite{McHardy2006}. Such a scaling relation has been shown to extend to BH XRB and weakly accreting NS systems \cite{Koerding2007} provided the frequency of the high frequency Lorentzians in these sources is physically related to the high frequency break seen in AGN. 
However, in AGN significant variability can be still measured at frequencies where BH XRBs variability appears to be largely suppressed (e.g. for AGN of $M_{BH}\sim10^{6-7}M_{\odot}$, high $F_{\rm var}$ can be measured at $\nu\gsim 10^{-4}$ Hz, which correspond to $\nu\gsim 10-100$ Hz for a BH XRB after scaling for the difference in BH mass). Such kind of discrepancies in comparing AGN and BH XRBs variability properties may be intrinsic or may arise as a consequence of compared energy band effects (due to the different disc temperatures, the same bandpass catches different parts of the accretion flow in AGN and BH XRBs) and accretion states \cite[e.g.][]{Done2005}.

AGN also show the same kind of X-ray lags as seen in hard-state BH XRBs. The low temporal frequencies are indeed dominated by hard lags \cite{DeMarco2013}, which however result poorly sampled in AGN due to the limited exposure time of observations (typically $\sim100-150$ ks, corresponding to a minimum temporal frequency of $\sim10^{-5}$ Hz) which do not allow the extent of the frequency range where such lags dominate to be properly constrained with more standard Fourier techniques (NGC 4051 is the only AGN where measurements of hard lags have been performed up to temporal frequencies of $\nu\sim10^{-7}$ Hz; \cite{Papadakis2019}). At the same time, X-ray reverberation lags, which were in fact first detected in AGN \cite{Fabian2009}, have now been detected in a large number of sources \cite{DeMarco2011,DeMarco2013,Kara2016}. X-ray reverberation lags (measured in the response of the soft excess, the FeK, and the Compton hump components) in AGN can be observed up to very high frequencies ($\gsim10^{-3}$ Hz), and map distances of the order of a few $R_g$-light crossing time. 
In radio quiet, unabsorbed AGN, these lags have been proved to scale with BH mass \cite{DeMarco2013,Kara2016}, suggesting that the same geometry (a disc, likely extending close to ISCO, plus a compact X-ray source) describes these systems. However, once extended down to the stellar-mass BH range, this relation predicts reverberation lags much shorter than actually measured in hard state BH XRBs, suggesting substantial differences in the geometry of the innermost accretion flow (possibly ascribable to different accretion regimes). 

Finally, while state transitions are routinely observed in BH XRBs, similar phenomena might occur also in AGN but on much longer time scales. To date, the best candidate-state transition events in AGN are represented by the ``changing-look'' phenomenon \cite{Noda2018}, although the latter is observed to occur on time scales of the order of tens of years, which is orders of magnitude faster than expected from a simple scaling of the observed time scales in BH XRBs for the corresponding difference in BH mass.

\section{Constraining the variability process}

The similarities discussed in the previous section suggest that the same underlying variability process may be at work in systems powered by accretion onto a compact object, independently of its mass, and highlight the need for identifying a common feature which could be used to constrain models of variability. 

The observed behaviour of BH XRBs during an outburst implies that the variability process is non-stationary \citep[e.g.][]{Vaughan2003} on time scales $\gsim$days, therefore resulting in dramatic changes of the PDS, both in shape and amplitude. The work by 
\cite{Uttley2001} demonstrated that this non-stationary behaviour extends to the shortest time scale variations, implying that the distribution of variability power over temporal frequencies cannot strictly be considered a defining characteristic of the underlying variability process. As a matter of fact a considerable number of proposed models for the X-ray variability of BH-accreting systems (e.g. shot-noise models, self-organized criticality models, propagating fluctuation models) can correctly reproduce the overall shape of the PDS once parameters are properly chosen, confirming the fact that this tool is insufficient to discriminate among models. However, \cite{Uttley2001} also noted that the absolute (not normalized) rms of the aperiodic broad band variability closely tracks the observed changes in flux, implying that the rms scales with the mean flux in a linear way (the source is more variable when brighter, or analogously, more intense long-term changes of flux coincide with more intense rapid variability), thus resulting in a constant fractional rms during most part of each accretion state (see Fig. \ref{fig:RID_PC}, left panel). 
This rms-flux relation holds on any time scale tested so far, e.g. the short time scale rms variability responds linearly to changes in flux occurring even on time scales of months. Moreover, such a relation is observed during all known accretion states in BH XRBs, though the parameters of the relation (slope and intercept) change from one state to the other \citep[e.g.][]{Heil2012}. Finally, rms-flux relations have been found in different accreting systems (e.g. NS \cite{Uttley2004}; ULXs \cite{Heil2010}; accreting white dwarfs \cite{Scaringi2012}), including AGN \citep[e.g.][]{Vaughan2003a,Uttley2001} (but see \cite{Alston2016}). 
All in all, this suggests that the rms-flux relation represents an universal feature of the aperiodic variability in accreting systems, thus key to constrain the underlying variability process.

From a mathematical point of view a rms-flux relation present on all time scales derives from the exponential transformation of an input linear light curve (i.e. $x(t)\approx \exp[l(t)]$, \cite{Uttley2005}). If the variability power in the input light curve is equally spread over many frequencies, for the central limit theorem the distribution of $\log[x(t)]$ is Gaussian, therefore the observed fluxes $x(t)$ have a log-normal distribution. From a physical point of view, this means that the locally produced variability is linear, but physical processes (e.g. diffusion) within the accretion flow mix the locally produced variations together in such a way (multiplicatively, \cite{Uttley2005}) that the resulting observed variability is non-linear and has a log-normal distribution. 
The skewness of this log-normal distribution depends on the fractional rms as $F_{\rm{var}}(F_{\rm{var}}^2 +3)$, meaning that sources with a larger $F_{\rm{var}}$ are more likely to occasionally show large amplitude flares, as indeed observed (e.g. in narrow line Seyfert 1 AGN).

The rms-flux relation is now considered as a fundamental feature that any model seeking to explain the aperiodic X-ray variability in BH-accreting systems has to reproduce. 
The rms-flux relation implies that the long time scale behaviour of the source modulates its shorter time scale variability, thus ruling out models that assume variations on different time scales to be independent of one another (e.g. independent X-ray flares in a corona).

\subsection{A word about models of X-ray variability}

Turbulence in the disc will naturally lead to stochastic perturbations of the viscosity parameter.  
However, whether any produced perturbation will result in significant variations of the observed flux depends on the time scales involved. Since the accretion process is diffusive in nature, perturbations generated on time scales longer than the diffusion time will propagate inwards, thus modulating variability produced in the inner, hard X-ray emitting regions. On the other hand perturbations lasting less than the diffusion time will be locally dampened, thus not contributing significantly to the overall X-ray variability \citep[e.g.][]{Lyubarskii1997,Churazov2001}. 

Nonetheless, while perturbations in the flow are naturally expected, their exact nature is not known. Several recent theoretical efforts have addressed this problem by studying long-duration, GRMHD simulations of optically thick and geometrically thin discs, as well as of optically thin and geometrically thick discs \citep[e.g.][]{Hogg2016,Bollimpalli2020}. 
Interestingly, these simulations consistently catch the development of sustained fluctuations in the mass accretion rate that diffuse down to small radii and are able to produce radially coherent variability, thus reproducing the observed rms-flux relation.

\subsection{A word about QPOs theoretical models}

Many models have been proposed for both LFQPOs and HFQPOs, some of which have been developed for years and largely tested against data, while others are little more than ideas mentioned in scattered papers. 
The only type of LF QPO that has a relatively widely accepted explanation  is the type-C QPO. These QPOs have been ascribed to the Lense-Thirring precession of particles within the accretion flow, a Relativistic effect that arises from the frame dragging occurring in the vicinity of spinning massive compact objects \cite{Ingram2009,Motta2018}. 
Despite the remarkably low number of detections of HFQPOs in BH XRBS, the majority of QPO theoretical models address the nature of such high-frequency features. This is most likely to be ascribed, as mentioned earlier, to the fact that their frequencies are relatively close to the orbital frequency of matter in the close vicinity of a BH. All models involve in some form the geodesic motion of matter around a massive object, modified by strong gravity as predicted by the General Theory of Relativity.
A detailed review of the several theoretical models for QPOs is beyond the scope of this work, and we refer the interested reader to \cite{Ingram2019} and references therein.

\section{Future perspectives}

The study of X-ray variability in BH systems is now largely recognised as an important milestone for a complete understanding of BH-accretion. 
As a result, proposed observing strategies with currently available X-ray satellites and detectors are increasingly more devoted to gather information from the observed X-ray variability in these sources.
In addition, incredible effort is being put by the broad community for the next generation of X-ray missions to be specifically designed to allow the potential of  X-ray timing to be fully exploited.
Successful application of timing and cross-spectral timing techniques requires the following ingredients: (i) high number of collected photons over the time scales of the accretion flow we want to probe; (ii) long exposures relative to the scales we want to probe; (iii) the possibility to observe very bright sources; (iv) high time resolution to allow sensitivity to the most rapid variability from the innermost accretion flow to be achieved. 
Thus, great progress is expected with the advent of X-ray detectors with large collecting area. The currently operative {\it NICER} payload \cite{Gendreau2016} and the Chinese Hard X-ray Modulation Telescope (HXMT) are already demonstrating how such high throughput facilities can contribute to improve our understanding of X-ray variability phenomena in BH-accreting systems. However, the greatest advancements will most likely come from the application of a ``multi-method'' analysis approach, whereby X-ray timing and cross-spectral timing techniques are used in concert with orthogonal analysis methods (e.g. polarimetry, \cite{Ingram2022}, or high resolution spectroscopy, \cite[e.g.][]{Barret2019}). This will be possible in the near future, thanks to the advent of the enhanced X-Ray Timing and Polarimetry (eXTP, \cite{DeRosa2019}) and the Advanced Telescope for High-ENergy Astrophysics (ATHENA, \cite{Nandra2013}). These observatories will combine large collecting area, good timing capabilities, and the possibility to perform simultaneous polarimetric (thanks to the Polarimetry Focusing Array PFA on eXTP) or spectroscopic (thanks to the Spectroscopic Focusing Array SFA on eXTP, and the X-ray Integral Field Unit X-IFU on ATHENA) measurements, which will undoubtedly make a huge step forward in our understanding of BH XRBs, and of accreting systems as a whole.

\vspace{0.3cm}



\section{Cross-References}
''Basics of Fourier Analysis for High-Energy Astronomy'', Belloni, T. M., Bhattacharya, D.

\bibliographystyle{model2-names-astronomy.bst}
\bibliography{bh_rev.bib}

\end{document}